\documentclass[aps,pra,twocolumn,showpacs,superscriptaddress]{revtex4-1}

\usepackage{amsmath,amssymb,amsfonts,bm,latexsym}
\usepackage[pdftex]{graphicx,color}
\usepackage{hyperref}
\hypersetup{
   colorlinks=true,        % false: boxed links; true: colored links
   linkcolor=blue,          % color of internal links (change box color with linkbordercolor)
   citecolor=blue,        % color of links to bibliography
  filecolor=magenta,      % color of file links
   urlcolor=blue           % color of external links
}

\newcommand{\erase}[1]{}

\newcommand{\bra}[1]{\langle #1|}
\newcommand{\ket}[1]{|#1 \rangle}
\newcommand{\bracket}[2]{\langle #1|#2 \rangle}

\newcommand{\ua}{\uparrow}
\newcommand{\da}{\downarrow}

\newcommand{\av}{{\bm{a}}}
\newcommand{\bv}{{\bm{b}}}
\newcommand{\cv}{{\bm{c}}}
\newcommand{\rv}{{\bm{r}}}
\newcommand{\kv}{{\bm{k}}}
\newcommand{\uv}{{\bm{u}}}
\newcommand{\vv}{{\bm{v}}}

\newcommand{\nv}{{\bm{n}}}

\newcommand{\hv}{\bm{h}}
\newcommand{\Hv}{\bm{H}}

\newcommand{\Sv}{\bm{S}}
\newcommand{\Tv}{\bm{T}}
\newcommand{\Jv}{\bm{J}}

\newcommand{\Dv}{\bm{D}}
\newcommand{\Dvb}{\bar{\bm{D}}}

\newcommand{\Kv}{\bm{K}}

\newcommand{\Mv}{\bm{M}}
\newcommand{\Nv}{\bm{N}}
\newcommand{\Gv}{\bm{G}}

\newcommand{\sigmav}{{\bm{\sigma}}}

\newcommand{\Jcal}{\mathcal{J}}
\newcommand{\Dcal}{\mathcal{D}}
\newcommand{\Scal}{\mathcal{S}}
\newcommand{\Dcalv}{\bm{\mathcal{D}}}
\newcommand{\Dcalvb}{\bar{\bm{\mathcal{D}}}}

\newcommand{\Ccal}{{\cal C}}
\newcommand{\Ical}{{\cal I}}

\newcommand{\dt}{\tilde{d}}
\newcommand{\at}{\tilde{a}}

\newcommand{\Scalt}{\tilde{\mathcal{S}}}
\newcommand{\Tt}{\tilde{T}}

\newcommand{\kinetic}{\mathrm{kin}}
\newcommand{\interact}{\mathrm{int}}
\newcommand{\BZ}{\mathrm{BZ}}

\newcommand{\Tesla}{\mathrm{T}}

\newcommand{\DM}{\mathrm{DM}}
\newcommand{\Ntri}{N_\mathrm{t}}

\newcommand{\cone}{\mathrm{c1}}

\allowdisplaybreaks[1]

\begin{document}

%%%%%%%%%%%%%%%%%%%%%%%%%%%%%%%%%%%%%%%%%%%%%%%%%
% Paper Information
%%%%%%%%%%%%%%%%%%%%%%%%%%%%%%%%%%%%%%%%%%%%%%%%%
\title{
Effects of Dzyaloshinskii-Moriya interactions in volborthite: \\
Magnetic orders and thermal Hall effect
}
\author{Shunsuke Furukawa}
\affiliation{Department of Physics, Keio University, 3-14-1 Hiyoshi, Kohoku-ku, Yokohama 223-8522, Japan}
%\affiliation{Department of Physics, University of Tokyo, 7-3-1 Hongo, Bunkyo-ku, Tokyo 113-0033, Japan}
\author{Tsutomu Momoi}
\affiliation{Condensed Matter Theory Laboratory, RIKEN, Wako, Saitama 351-0198, Japan}
\affiliation{RIKEN Center for Emergent Matter Science (CEMS), Wako, Saitama 351-0198, Japan}
\date{\today}
%\keywords{}

% 03.75.Hh Static properties of condensates; thermodynamical, statistical, and structural properties
% 03.75.Lm Tunneling, Josephson effect, Bose-Einstein condensates in periodic potentials, solitons, vortices, and topological excitations
% (see also 74.50.+r Tunneling phenomena; Josephson effects in superconductivity)
% 03.75.Mn Multicomponent condensates; spinor condensates
% 05.30.Jp Boson systems
% (for static and dynamic properties of Bose-Einstein condensates, see 03.75.Hh and 03.75.Kk; see also 67.10.Ba Boson degeneracy in quantum fluids)
% 67.85.-d	Ultracold gases, trapped gases (see also 03.75.-b Matter waves in quantum mechanics)
% 67.85.Fg	Multicomponent condensates; spinor condensates
% 72.25.-b Spin polarized transport
% 73.43.-f Quantum Hall effects
% 73.43.Cd Theory and modeling
% 73.43.Nq Quantum phase transitions
%85.75.-d Magnetoelectronics; spintronics: devices exploiting spin polarized transport or integrated magnetic fields

%%%%%%%%%%%%%%%%%%%%%%%%%%%%%%%%%%%%%%%%%%%%%%%%%
% Abstract
%%%%%%%%%%%%%%%%%%%%%%%%%%%%%%%%%%%%%%%%%%%%%%%%%

\begin{abstract}
Volborthite offers an interesting example of a highly frustrated quantum magnet
in which ferromagnetic and antiferromagnetic interactions compete on anisotropic kagome lattices.
A recent density functional theory calculation has provided a magnetic model based on coupled trimers,
which is consistent with a broad $\frac13$-magnetization plateau observed experimentally.
Here we study the effects of Dzyaloshinskii-Moriya (DM) interactions in volborthite.
We derive an effective model in which pseudospin-$\frac12$ moments emerging on trimers
form a network of an anisotropic triangular lattice.
Using the effective model, we show that for a magnetic field perpendicular to the kagome layer,
magnon excitations from the $\frac13$-plateau
feel a Berry curvature due to the DM interactions, giving rise to a thermal Hall effect.
Our magnon Bose gas theory can explain qualitative features of the magnetization and
the thermal Hall conductivity measured experimentally.
A further quantitative comparison with experiment poses constraints on the coupling constants in the effective model, promoting a quasi-one-dimensional picture.
Based on this picture, we analyze low-temperature magnetic phase diagrams using effective field theory,
and point out their crucial dependence on the field direction.
\end{abstract}

\maketitle

%%%%%%%%%%%%%%%%%%%%%%%%%%%%%%%%%%%%%%%%%%%%%%%%%
% Main text
%%%%%%%%%%%%%%%%%%%%%%%%%%%%%%%%%%%%%%%%%%%%%%%%%

%%%%%%%%%%%%%%%%%%%%%%%%%%%%%%%%%%%%%%%%%%%%%%%%%
\section{Introduction}\label{sec:intro}
%%%%%%%%%%%%%%%%%%%%%%%%%%%%%%%%%%%%%%%%%%%%%%%%%

% [ Frustration ]--------------------

The last two decades have witnessed increasing interest in highly frustrated quantum magnetism,
boosted by the development in theoretical concepts and simulation methods as well as a huge variety of material realizations \cite{Lacroix11,Diep05,Vasiliev18}.
%\tm{$\leftarrow$ slightly unreadable because of two ands}
On one hand, such magnets do not easily find energetically stable states, which results in a wide range of behavior under perturbations.
On the other hand, they offer an attractive possibility of a quantum spin liquid (QSL),
which evades any ordering in terms of a conventional order parameter down to zero temperature \cite{Balents10,Savary16,Knolle19}.
Antiferromagnets on triangular and kagome lattices are typical examples of geometrical frustration in two dimensions.
Numerical studies on the spin-$\frac12$ kagome antiferromagnetic Heisenberg model have provided indications of a QSL ground state
with either gapped \cite{Yan11,Depenbrock12,Jiang12,Mei17} or gapless \cite{Iqbal14,Liao17,He17,Jiang19} low-energy excitations
although its precise nature is still under active debate.
The enigmatic nature of the kagome antiferromagnet has stimulated experimental studies on a number of copper minerals \cite{Inosov19}
such as herbertsmithite \cite{Helton07,Norman16}, volborthite \cite{Hiroi01,Hiroi19}, and vesignieite \cite{Okamoto09}, which host layers of spin-$\frac12$ moments arranged in a kagome pattern.
Thermodynamic and neutron scattering measurements for herbertsmithite
have identified the gapless QSL behavior with fractionalized excitations \cite{Helton07,Han12}
while the NMR experiment has detected a small intrinsic excitation gap \cite{FuImai15}.
As for the scenario of rich behavior due to frustration,
the spatially anisotropic triangular antiferromagnet Cs$_2$CuCl$_4$ offers a particularly interesting example.
Experiments have revealed continuum of excitations indicative of spin fractionalization \cite{Coldea01,Coldea03},
and rich magnetic phase diagrams that crucially depend on the field direction \cite{Coldea01,Tokiwa06}.
A theoretical understanding of these results has been made by viewing the system as weakly coupled Heisenberg chains \cite{Starykh07,Starykh10,Kohno07,Kohno09,Starykh15},
and thereby an extreme sensitivity to weak magnetic anisotropy and inter-layer couplings has been pointed out \cite{Starykh10}.

%Frustrated quantum spin systems have been a subject of intensive theoretical and experimental research \cite{Vasiliev18}.
%There has been an ever growing interest in frustration effects in quantum spin systems
%owing to the discovery and the synthesis of a variety of magnetic materials with triangle-based networks of magnetic ions \cite{Lacroix11,Diep05}.
%
%Intuitively, QSLs are results of the fact that a variety of spin configurations have nearly degenerate energies
%and form a quantum mechanical superposition owing to quantum fluctuations.
%This in turn suggests that frustrated magnets
%Frustrated quantum magnets are also attractive platform for observing a variety of symmetry breaking because of their high sensitivity to perturbations.

\begin{figure}[b]
\begin{center}\includegraphics[width=0.45\textwidth]{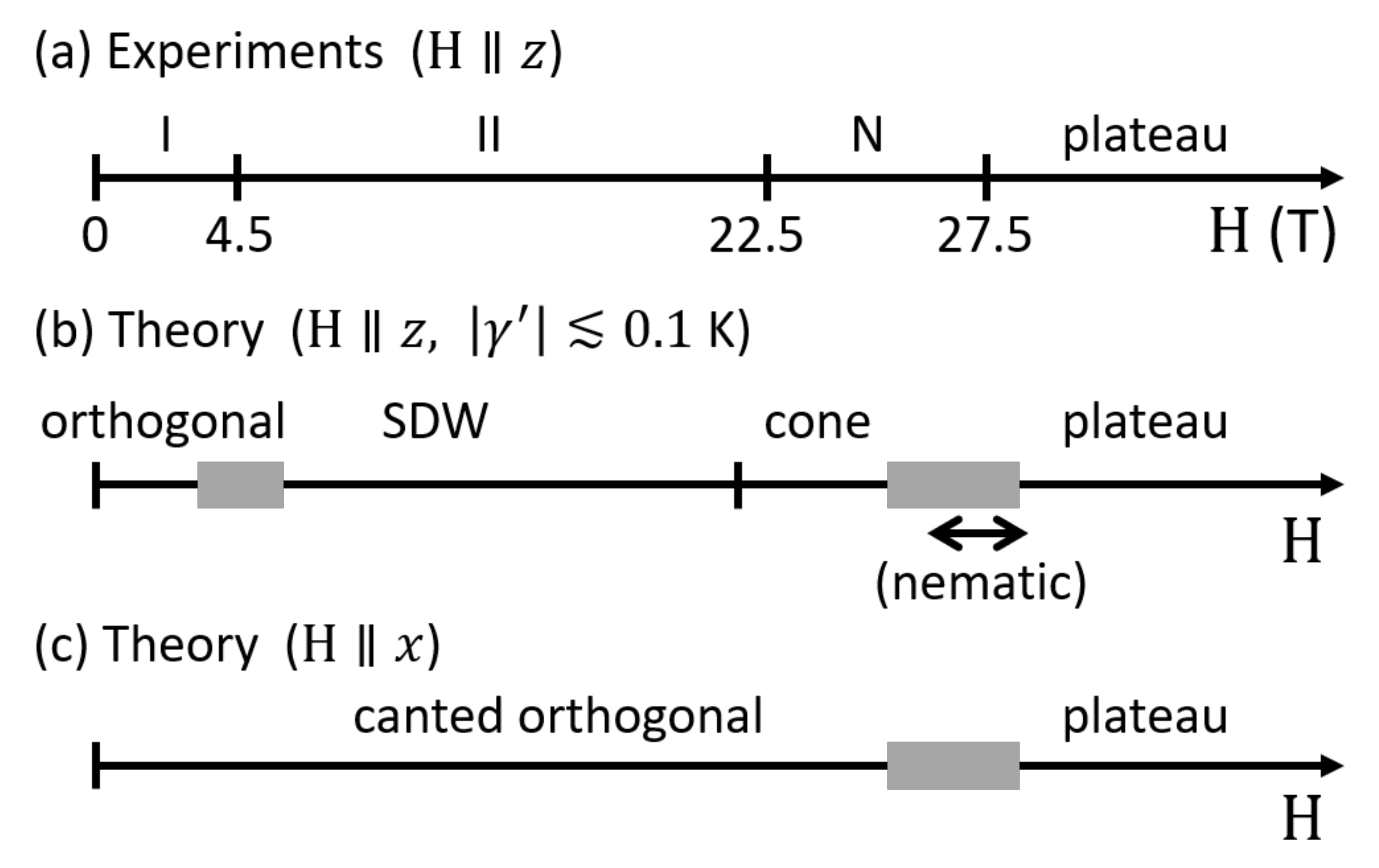}\end{center}
\caption{\label{fig:phasemag} %(Color online)
Schematic magnetic phase diagrams at low temperatures based on experiments (single crystals) and theory.
See Fig.\ \ref{fig:volbeffDM} for the definitions of directions.
In (a) the experimental results for $\Hv\parallel z$,
Phase II shows double-horn NMR spectra indicative of an incommensurate SDW order \cite{Ishikawa15,YoshidaM17}.
Phase N can be divided into two phases, N$_1$ and N$_2$ \cite{Kohama19}.
The theoretical results in (b) and (c) based on effective field theory for a quasi-one-dimensional regime
exhibit a marked sensitivity to the field direction due to the DM interactions.
In (b), we assume that the magnitude of the strongly relevant interaction $\gamma'$
between the second-neighbor chains [defined in Eq.\ \eqref{eq:gammap} later] is suppressed ($|\gamma'|\lesssim 0.1$ K).
The nature of the obtained phases are described in detail in Sec.\ \ref{sec:fieldtheory}.
Shaded areas indicate the regimes where the field-theoretical approach is less effective (owing to spatially oscillating interactions or a vanishing velocity).
However, the analysis of the multi-magnon spectra indicates that the bond nematic state due to condensation of bimagnons
appears just below the plateau for a certain parameter range \cite{Janson16}.
}
\end{figure}

% [ Volborthite ]--------------------
In this context, volborthite Cu$_3$V$_2$O$_7$(OH)$_2\cdot$2H$_2$O is a fascinating material for which a wealth of field-induced phenomena have been observed
in powder \cite{YoshidaH09,YoshidaM09,YoshidaM11,YoshidaM12} and single-crystal \cite{Ishikawa15,YoshidaM17,Kohama19,Watanabe16,Yamashita20,Nakamura18,Ikeda19} samples.
The material was originally considered as a candidate for a spin-$\frac12$ kagome antiferromagnet \cite{Hiroi01},
and a sign of strong frustration has been found in the low magnetic transition temperature around $1$ K
in comparison with the Curie-Weiss temperature $-155$ K \cite{Fukaya03,Bert05,YoshidaM09,YoshidaH12}. %Yamashita10,Nilsen11,
X-ray diffraction measurements for single crystals have, however, suggested highly anisotropic arrangements of magnetically active orbitals at the crystallographically distinct Cu sites,
indicating a strong spatial anisotropy in magnetic interactions \cite{YoshidaH12,Ishikawa15,Hiroi19}.
Furthermore, magnetization measurements for single crystals have revealed a wide $\frac13$-magnetization plateau which starts at $H=27.5$ T \cite{Ishikawa15};
according to recent Faraday rotation measurements, this plateau continues as high as 100 T or even above 160 T, depending on the sample setting condition \cite{Nakamura18}.
Such an extremely wide $\frac13$-plateau is in sharp contrast with the relatively narrow plateau in the kagome antiferromagnetic model \cite{Nishimoto13,Capponi13,Schulenburg02}.
Below this plateau, NMR measurements have identified three distinct phases \cite{Ishikawa15,YoshidaM17}, as schematically shown in Fig.\ \ref{fig:phasemag}(a).
Phase II shows double-horn NMR spectra indicative of an incommensurate spin-density-wave (SDW) order
while indications of bimagnon condensation upon entering Phase N from the plateau have been found in Refs.~\cite{YoshidaM17,Kohama19}.
Thermal measurements \cite{Kohama19} have indicated that Phase N is divided into two phases, N$_1$ and N$_2$.

%############################
%############################

\begin{figure*}
\begin{center}
\includegraphics[width=\textwidth]{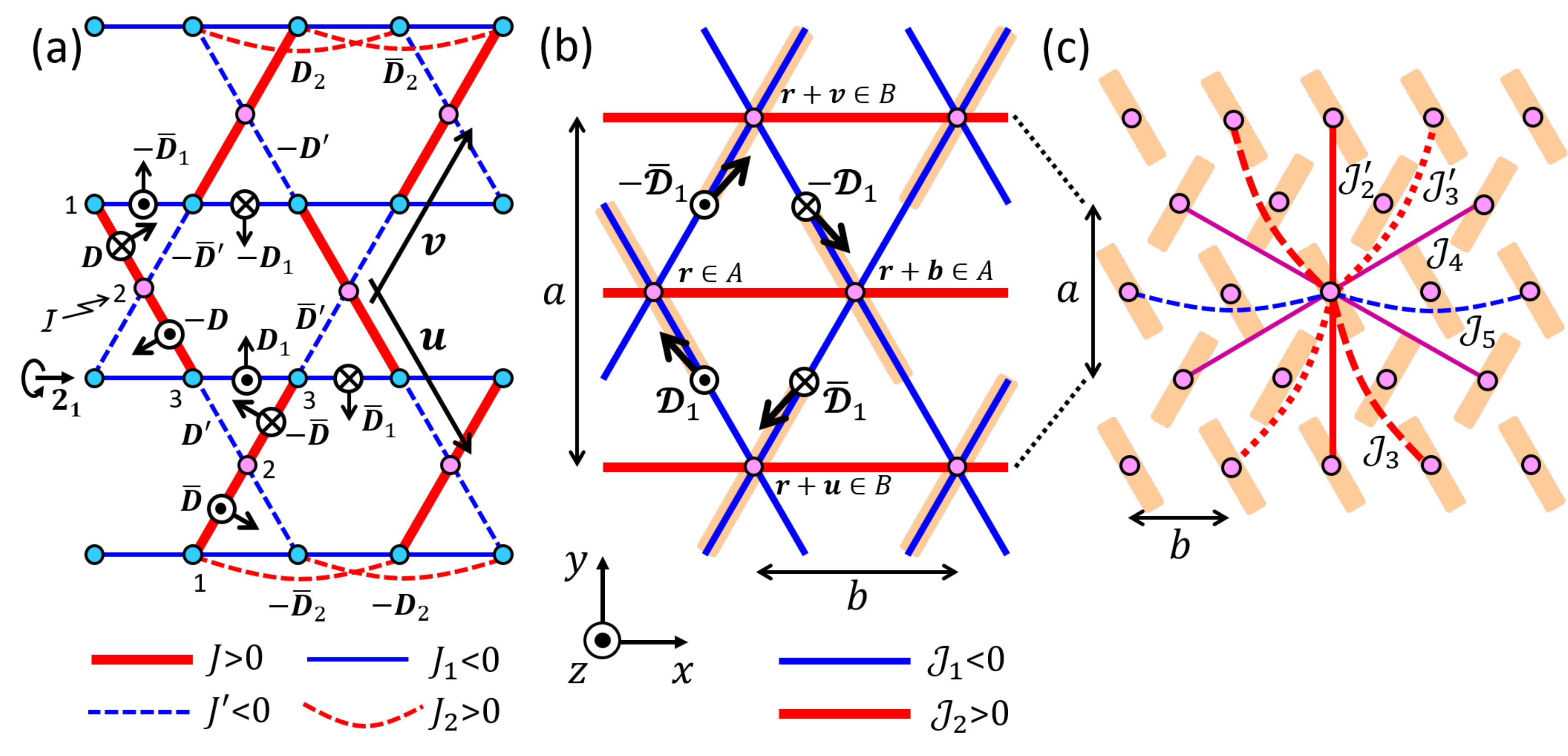}
\end{center}
\caption{\label{fig:volbeffDM}
(Color online)
Schematic diagrams of (a) the microscopic spin-$\frac12$ model \eqref{eq:H_volb}
and (b,c) the effective pseudospin-$\frac12$ model \eqref{eq:Heff2} for a single magnetic layer of volborthite.
Here, (b) shows only the nearest-neighbor interactions which appear in the first-order strong-coupling expansion, and
(c) shows further-neighbor interactions which appear in the second-order expansion.
Some of the interactions are shown only for representative bonds for brevity.
See Eqs.~\eqref{eq:J_volb}, \eqref{eq:DvDv1}, and \eqref{eq:Dcalv1} and Table \ref{table:estimates} for the estimates of the (effective) coupling constants based on DFT+$U$ \cite{Janson16}.
For simplicity, we work with the approximate lattice in which all the spins on the layer reside in the $xy$ plane,
and the central sites of neighboring trimers are connected by the vectors $\uv=(b,-a,0)/2$, $\vv=(b,a,0)/2$, and $\bv=(b,0,0)$.
In (a), three spins on each trimer are labeled by $j=1,2,3$ in ascending order from left to right.
In defining DM vectors, every bond is oriented from left to right, and an arrow (a dot or a cross) indicates the in-plane (out-of-plane) component.
The (effective) DM vectors are subject to constraints due to crystal symmetries such as
the inversion ${\cal I}$ about the center of each trimer and the two-fold screw ($2_1$) axis along each $J_1$-$J_2$ chain (see Appendix \ref{app:DM}).
A bar on a DM vector indicates $\pi$ rotation around the $x$ axis.
%The $x$ and $y$ directions correspond to the crystallographic $b$ and $a$ ($c$) directions in the notation of Ref.~\cite{Ishikawa15} (Ref.~\cite{Janson16}).
}
\end{figure*}

% [ Coupled-trimer model ]--------------------
In Ref.\ \cite{Janson16}, density functional theory (DFT) calculations have been performed
on the basis of the single-crystal structural data \cite{Ishikawa15} to determine the microscopic spin model of volborthite.
Using DFT+$U$, four leading exchanges have been identified, as schematically shown in Fig.\ \ref{fig:volbeffDM}(a):
antiferromagnetic $J$ and $J_2$ as well as ferromagnetic $J'$ and $J_1$ with a distinctive hierarchy $J>|J_1|>J_2,|J'|$.
The dominance of the $J$ coupling naturally leads to a coupled-trimer picture:
on each trimer formed by the $J$ coupling, the spin states are restricted to the lowest-energy doublet,
which can be viewed as pseudospin-$\frac12$ states, at zero field
and such pseudospins interact with each other through the inter-trimer couplings.
This sharply contrasts with the coupled frustrated chain model with $J=J'$ \cite{Janson10}, which was obtained previously based on powder structural data.
In the coupled-trimer model, the $\frac13$-plateau state can be interpreted as a product of polarized trimers, and a wide plateau extending to $H=225$ T has been predicted.
By means of a strong-coupling expansion, an effective model has been derived
for the pseudospin-$\frac12$ degrees of freedom living on an anisotropic triangular lattice, as shown in Fig.\ \ref{fig:volbeffDM}(b,c).
This model shows a tendency towards condensation of magnon bound states preceding the plateau,
indicating the emergence of a bond nematic order \cite{Momoi05,Shannon06,Kecke07,Hikihara08,Sudan09,Balents16}.
This can provide a scenario for Phase N (or N$_2$) observed experimentally \footnote{We note that the analysis of the coupled frustrated chain model
has led to yet another scenario, a chiral liquid \cite{Parker17}.}.
The coupled-trimer model has also stimulated theoretical studies of possible QSLs above magnetic ordering temperatures \cite{ChernHwang17,ChernSchaffer17}.

%The obtained model provides a unified description of various intriguing properties of this compound

%############################
%############################

% [ Possible magnetic anisotropy ]--------------------
The theoretical analyses of Ref.\ \cite{Janson16} have mostly been based on isotropic Heisenberg interactions.
Although Ref.\ \cite{Janson16} has also given estimates of the Dzyaloshinskii-Moriya (DM) interactions for the leading couplings $J$ and $J_1$,
their effects on the magnetic properties have not been analyzed in detail.
Given a high sensitivity of frustrated magnets,
the DM interactions can significantly influence the low-temperature magnetic orderings.
Furthermore, these interactions can give rise to nontrivial transport properties such as a magnon thermal Hall effect \cite{Onose10,Matsumoto11_PRL, Matsumoto11_PRB, Murakami17}.
In fact, a thermal Hall effect has been observed in volborthite in magnetic fields up to 15 T perpendicular to the magnetic layer \cite{Watanabe16,Yamashita20}.
The dependence of the observed transverse thermal response on the magnetic field indicates that the effect is due to spin excitations.
Interestingly, the effect has been observed even above the magnetic transition temperatures,
where the system may be described as a cooperative paramagnet.

%the effect has been observed even above
%Interestingly, the thermal Hall conductivity has been found to change its sign when the temperature is lowered
%from the high-temperature disordered phase to the low-temperature SDW phase.

% [ This paper ]--------------------
In this paper, we theoretically study the effects of DM interactions in volborthite on the basis of the coupled-trimer model of  Ref.\ \cite{Janson16}.
By incorporating the effects of the DM interactions,
we derive an effective pseudospin-$\frac12$ model on an anisotropic triangular lattice.
In the resulting model, the magnetic anisotropy is characterized
by a single effective DM vector (see $\Dcalv_1^z$ in Fig.\ \ref{fig:volbeffDM}), which leads to a significant simplification of the analysis.
We show that for a magnetic field perpendicular to the kagome layer,
magnon excitations from the $\frac13$-plateau (fully polarized pseudosopins) feel a nonzero Berry curvature due to the effective DM interaction,
leading to a thermal Hall effect.
This effect disappears when the field is changed to the direction of the screw axis (the $x$ direction of Fig.\ \ref{fig:volbeffDM}), reflecting a symmetry.
Although our analysis of the thermal Hall effect is based on the spin wave picture valid just below the plateau,
we expect that the magnitude and qualitative features of the thermal Hall conductivity do not change abruptly
as we lower the magnetic field to the regime $|H|\lesssim 15$ T investigated experimentally \cite{Watanabe16,Yamashita20}.
Comparison of the theory with experimental data of the magnetization process and the thermal Hall conductivity
poses constraints on the coupling constants in the effective model, promoting a quasi-one-dimensional picture.
Based on this picture, we further analyze magnetic orders at low temperatures using effective field theory (in close parallel with the theory for Cs$_2$CuCl$_4$ \cite{Starykh07,Starykh10}),
and point out that the magnetic phase diagram can sensitively depend on the field direction owing to the DM interaction, as shown in Fig.\ \ref{fig:phasemag}(b,c).
We thus find a striking similarity to the physics of Cs$_2$CuCl$_4$.

% [ Organization of the paper ]--------------------
The rest of this paper is organized as follows.
In Sec.~\ref{sec:model}, we describe the microscopic spin-$\frac12$ model obtained by DFT+$U$ for volborthite in Ref.\ \cite{Janson16}.
We then perform a strong-coupling expansion to derive an effective pseudospin-$\frac12$ model on an anisotropic triangular lattice.
In Sec.~\ref{sec:spwv}, we perform a spin wave analysis starting from the $\frac13$-plateau state,
and analyze the field- and temperature-dependences of the magnetization and the thermal Hall conductivity.
In Sec.~\ref{sec:fieldtheory}, we analyze low-temperature magnetic orders in light of an effective field theory for a quasi-one-dimensional regime.
In Sec.~\ref{sec:summary}, we present a summary and an outlook for future studies.

%%%%%%%%%%%%%%%%%%%%%%%%%%%%%%%%%%%%%%%%%%%%%%%%%
\section{Model} \label{sec:model}
%%%%%%%%%%%%%%%%%%%%%%%%%%%%%%%%%%%%%%%%%%%%%%%%%

In this section, we describe the microscopic spin-$\frac12$ model obtained by DFT+$U$ \cite{Janson16},
and then perform a strong-coupling expansion to derive an effective pseudospin-$\frac12$ model on an anisotropic triangular lattice.

Before presenting the theoretical models, let us briefly review the low-temperature crystal structure of volborthite.
Below $155$ K, volborthite shows the P$2_1/a$ structure (space group No.\ 14) with the lattice constants \cite{Hiroi19,Ishikawa15}
\begin{equation}
\begin{split}
 &a=10.6489(1)~\AA,~b=5.8415(1)~\AA,\\
 &c=14.4100(1)~\AA,~\beta=95.586(1)^\circ.
\end{split}
\end{equation}
We introduce the Cartesian coordinate $(x,y,z)$ in such a way that the $x$ and $y$ axes are along the crystallographic $b$ and $a$ directions, respectively.
In this coordinate, the primitive vectors of the lattice are represented as
\begin{equation}\label{eq:abcbeta}
 \bv=(b,0,0),~\av=(0,a,0),~\cv=c(0,\cos\beta,-\sin\beta).
\end{equation}
A single kagome layer extends in the $xy$ or $ab$ plane as shown in Fig.\ \ref{fig:volbeffDM},
and two kinds of kagome layers are stacked alternately in the $c$ direction.
The layer dependence of exchange couplings is estimated to be small \cite{Janson16}, and neglected in the subsequent analysis.

%The following symmetries of the crystal are relevant to the analyses in this paper.
%Firstly, there is an inversion center $I$ at the center of each trimer.
%Secondly, a two-fold screw ($2_1$) axis runs along the $x$ (or $b$) direction, as shown in Fig.\ \ref{fig:volbeffDM}.

%================================================
\subsection{Microscopic spin-$\frac12$ model}
%================================================

The microscopic spin-$\frac12$ model determined by DFT+$U$ \cite{Janson16} is shown in Fig.\ \ref{fig:volbeffDM}(a).
We label each trimer by the location $\rv$ of its central site.
Let $\Sv_{\rv,j}~(j=1,2,3)$ be the three spin-$\frac12$ operators on the trimer.
There are two types of trimers, $A$ and $B$;
the sets of the central sites of those trimers are denoted by $A$ and $B$ as well.
%The crystal symmetry imposes constraints on the relative signs of DM vectors as discussed in Appendix \ref{app:DM}.

% [ Hamiltonian ]--------------------------------------
The Hamiltonian is given by
\begin{equation}\label{eq:H_volb}
 H = \sum_{\rv\in A\cup B} \left( H_\rv + H_{\rv,\rv+\uv} + H_{\rv,\rv+\vv} + H_{\rv,\rv+\bv} \right),
\end{equation}
where the vectors $\uv$, $\vv$, and $\bv$ connect between neighboring trimers as shown in Fig.\ \ref{fig:volbeffDM}.
For a trimer at $\rv\in X=A,B$, the intra-trimer interactions are given by
\begin{equation*}
\begin{split}
 H_\rv &= J (\Sv_{\rv,1}\cdot\Sv_{\rv,2}+\Sv_{\rv,2}\cdot\Sv_{\rv,3})\\
 &+ \Dv_X \cdot (\Sv_{\rv,1}\times\Sv_{\rv,2}-\Sv_{\rv,2}\times\Sv_{\rv,3})
 -\hv\cdot \sum_{j=1}^3 \Sv_{\rv,j}
\end{split}
\end{equation*}
with $\Dv_A=\Dv$, $\Dv_B=\Dvb$, and $\hv=-g\mu_B\Hv/k_B$.
Here, a Zeeman energy in an external magnetic field $\Hv$ has been included.
Throughout this paper, a bar on a vector indicates $\pi$ rotation around the $x$ axis, i.e., $\Dvb=(D^x,-D^y,-D^z)$.
The inter-trimer interactions along $\uv$ and $\vv$ are given for $\rv\in A$ by
\begin{equation*}
\begin{split}
 H_{\rv,\rv+\uv}
 =& J_1 \Sv_{\rv,3}\cdot\Sv_{\rv+\uv,3}+J' \Sv_{\rv,3}\cdot\Sv_{\rv+\uv,2}\\
 &+\Dv_1\cdot (\Sv_{\rv,3}\times\Sv_{\rv+\uv,3})+\Dv'\cdot (\Sv_{\rv,3}\times\Sv_{\rv+\uv,2}),\\
 H_{\rv,\rv+\vv}
 =& J_1 \Sv_{\rv,1}\cdot\Sv_{\rv+\vv,1}+J' \Sv_{\rv,2}\cdot\Sv_{\rv+\vv,1}\\
 &-\Dvb_1\cdot (\Sv_{\rv,1}\times\Sv_{\rv+\vv,1})-\Dvb'\cdot (\Sv_{\rv,2}\times\Sv_{\rv+\vv,1}),
\end{split}
\end{equation*}
and for $\rv\in B$ by
\begin{equation*}
\begin{split}
 H_{\rv,\rv+\uv}
 =& J_1 \Sv_{\rv,1}\cdot\Sv_{\rv+\uv,1}+J' \Sv_{\rv,2}\cdot\Sv_{\rv+\uv,1}\\
 &-\Dv_1\cdot (\Sv_{\rv,1}\times\Sv_{\rv+\uv,1})-\Dv'\cdot (\Sv_{\rv,2}\times\Sv_{\rv+\uv,1}),\\
 H_{\rv,\rv+\vv}
 =& J_1 \Sv_{\rv,3}\cdot\Sv_{\rv+\vv,3}+J' \Sv_{\rv,3}\cdot\Sv_{\rv+\vv,2}\\
 &+\Dvb_1 \cdot (\Sv_{\rv,3}\times\Sv_{\rv+\vv,3})+\Dvb'\cdot (\Sv_{\rv,3}\times\Sv_{\rv+\vv,2}).
\end{split}
\end{equation*}
Furthermore, the inter-trimer interactions along $\bv$ are given for $\rv\in X=A,B$ by
\begin{equation*}
\begin{split}
 H_{\rv,\rv+\bv}
 =& J_2 (\Sv_{\rv,1}\cdot\Sv_{\rv+\bv,1}+\Sv_{\rv,3}\cdot\Sv_{\rv+\bv,3})\\
 & -\Dv_{2X}\cdot (\Sv_{\rv,1}\times\Sv_{\rv+\bv,1}-\Sv_{\rv,3}\times\Sv_{\rv+\bv,3})
\end{split}
\end{equation*}
with $\Dv_{2A}=\Dv_2$ and $\Dv_{2B}=\Dvb_2$.

% [ Coupling constants ]--------------------------------------
Based on DFT+$U$ and the fit with the magnetic susceptibility data,
the Heisenberg exchange couplings and the Land\'e $g$-factor have been estimated as
\begin{equation}\label{eq:J_volb}
\begin{split}
 &J:J':J_1:J_2=1:-0.2:-0.5:0.2,\\
 &J=252~\mathrm{K},~g=2.151.
\end{split}
\end{equation}
Furthermore, the DM interactions on the $J$ and $J_1$ bonds are estimated
(after appropriate rescaling as noted in Ref.~\cite{Janson16}) as
\begin{equation}\label{eq:DvDv1}
\begin{split}
 \Dv&=(D^x,D^y,D^z)=(12,7,-17)~\mathrm{K},\\
 \Dv_1&=(D_1^x,D_1^y,D_1^z)=(0,3,8)~\mathrm{K}.
\end{split}
\end{equation}
The DM vectors $\Dv'$ and $\Dv_2$ on the $J'$ and $J_2$ bonds have not been estimated.
We note that the presence of the ferromagnetic $J_1<0$ coupling is consistent
with negative magnetostriction along the $x$ axis observed recently \cite{Ikeda19}.

For the direction of the external field $\Hv$, we mainly consider the following two cases:
\begin{subequations}
\begin{align}
 \text{(i)}~&\Hv=(0,0,-H)\parallel -\hat{z},~\text{i.e.,}~\hv=(0,0,h)\parallel \hat{z}; \label{eq:H_z}\\
 \text{(ii)}~&\Hv=(-H,0,0)\parallel -\hat{x},~\text{i.e.,}~\hv=(h,0,0)\parallel \hat{x} \label{eq:H_x}
\end{align}
\end{subequations}
with $h=g\mu_B H/k_B>0$.
Here, $\Hv$ and $\hv$ point in mutually opposite directions
because of the negative $g$-factor $-g<0$ of an electron.
Experiments have mainly been performed for $\Hv \parallel \pm \hat{z}$.

%================================================
\subsection{Isolated trimer}
%================================================

% [ SU(2) case ]--------------------------------------
We first consider the Hamiltonian for an isolated trimer of type $A$, which is given by
\begin{equation}\label{eq:H_trimer}
\begin{split}
 H_\mathrm{trimer}=
 &J(\Sv_1\cdot\Sv_2+\Sv_2\cdot\Sv_3) \\
 &+ \Dv\cdot (\Sv_1\times\Sv_2-\Sv_2\times\Sv_3)
 -\hv\cdot \sum_{j=1}^3 \Sv_j.
\end{split}
\end{equation}
We note that the Hamiltonian for a trimer of type $B$ can be treated by replacing $\Dv$ by $\Dvb$ in the above.
When $\Dv=\bm{0}$ and ${\hv}=\bm{0}$, because of the SU(2) symmetry and the inversion symmetry around the site $2$,
the eigenstates of Eq.~\eqref{eq:H_trimer} are classified into the quadruplet $\{\ket{q_\mu}\}$,
the even-parity doublet $\{\ket{d_\mu}\}$, and the odd-parity doublet $\{\ket{d'_\mu}\}$,
where $\mu$ is the eigenvalue of $\sum_{j=1}^3 S_j^z$.
Their wave functions are given by
\begin{align}
\ket{q_{+\frac32}} &= \ket{\ua\ua\ua},\notag\\
\ket{q_{+\frac12}} &= \frac1{\sqrt{3}} S^- \ket{q_{+\frac32}} = \frac{1}{\sqrt{3}} \left( \ket{\ua\ua\da}+\ket{\da\ua\ua}+\ket{\ua\da\ua}\right),\notag\\
\ket{q_{-\frac12}} &= \frac1{2} S^- \ket{q_{+\frac12}} = \frac{1}{\sqrt{3}} \left( \ket{\da\da\ua}+\ket{\ua\da\da}+\ket{\da\ua\da}\right),\notag\\
\ket{q_{-\frac32}} &= \frac1{\sqrt{3}} S^- \ket{q_{-\frac12}}=\ket{\da\da\da},\notag\\
\ket{d_{+\frac12}} &= \frac{1}{\sqrt{6}} \left( \ket{\ua\ua\da}+\ket{\da\ua\ua}-2\ket{\ua\da\ua}\right),\notag\\
\ket{d_{-\frac12}} &= S^-\ket{d_{+\frac12}} =-\frac{1}{\sqrt{6}} \left( \ket{\da\da\ua}+\ket{\ua\da\da}-2\ket{\da\ua\da}\right),\notag\\
\ket{d'_{+\frac12}} &= \frac{1}{\sqrt{2}} \left( \ket{\ua\ua\da}-\ket{\da\ua\ua}\right),\notag\\
\ket{d'_{-\frac12}} &= S^-\ket{d'_{+\frac12}} =-\frac{1}{\sqrt{2}} \left( \ket{\da\da\ua}-\ket{\ua\da\da}\right),
\label{eq:eig_trimer}
\end{align}
where $S^-=\sum_{j=1}^3(S_j^x-i S_j^y)$.
The Zeeman term in Eq.~\eqref{eq:H_trimer} commutes with the Heisenberg terms,
and only shifts the eigenenergies by $-h\mu$ when $\hv=(0,0,h)$.
The eigenenergies in this case are calculated as
\begin{equation}\label{eq:eigener_trimer}
\begin{split}
 \ket{q_\mu}:&~\frac12 J-h\mu;\\
 \ket{d_\mu}:&~-J-h\mu;\\
 \ket{d_\mu'}:&~-h\mu.
\end{split}
\end{equation}
At $h=0$, the ground states are the even-parity doublet $\ket{d_{\pm\frac12}}$; they are split for $h\ne 0$.
For low temperatures and low fields ($T,|h|\ll J$), it is therefore legitimate to focus on
the pseudospin-$\frac12$ subspace $\mathrm{Span}\{\ket{d_{\pm\frac12}}\}$, as has been done in Ref.~\cite{Janson16}.

% [ With DM ]--------------------------------------
When $\Dv=(D^x,D^y,D^z)\ne\bm{0}$, however, the even-parity doublet $\{ \ket{d_{\pm\frac12}} \}$ is modified.
By assuming $|\Dv|\ll J$, the new eigenstates are given to first order in $\Dv$ by
\begin{equation}\label{eq:dt_trimer}
 \ket{\dt_{\pm \frac12}} = \ket{d_{\pm\frac12}} \pm \frac{iD^\mp}{\sqrt{6}J} \ket{q_{\pm\frac32}}
 -\frac{i\sqrt{2}D^z}{3J} \ket{q_{\pm\frac12}} \mp \frac{iD^\pm}{3\sqrt{2}J} \ket{q_{\mp\frac12}},
\end{equation}
where $D^\pm:=D^x\pm iD^y$.
The presence of $\hv=(0,0,h)\ne\bm{0}$ does not modify $\ket{\dt_{\pm \frac12}}$ in Eq.~\eqref{eq:dt_trimer} to first order in $h$
because the Zeeman term does not mix $\{\ket{d_\mu}\}$ with $\{\ket{q_\mu}\}$ or $\{\ket{d'_\mu}\}$
(however, it does modify the eigenenergies as in Eq.~\eqref{eq:eigener_trimer}).

%================================================
\subsection{Effective pseudospin-$\frac12$ model}
%================================================

% [ Local basis ]--------------------------------------
We adopt the lowest-energy doublet $\{\ket{\dt_{\pm \frac12}}_\rv\}$ at $\hv=\bm{0}$ as the local basis on each trimer $\rv\in A\cup B$,
and perform a strong coupling expansion \cite{Mila11} to derive the effective Hamiltonian
(see Refs.\ \cite{Tonegawa00,Honecker01} for related calculations for distorted diamond chains).
Using this doublet, we introduce a pseudospin-$\frac12$ operator
\begin{equation}
 \Tv_\rv=\frac12 \left( \ket{\dt_{+\frac12}}_\rv , \ket{\dt_{-\frac12}}_\rv  \right) \sigmav
 \begin{pmatrix} {}_\rv\bra{\dt_{+\frac12}} \\ {}_\rv\bra{\dt_{-\frac12}}  \end{pmatrix},
\end{equation}
where $\sigmav=(\sigma^x,\sigma^y,\sigma^z)$ are Pauli matrices.
We also introduce a local projection operator $P_\rv$ and a global one $P$ as
\begin{equation}
 P_\rv=\ket{\dt_{+\frac12}}_\rv {}_\rv\bra{\dt_{+\frac12}} + \ket{\dt_{-\frac12}}_\rv {}_\rv\bra{\dt_{-\frac12}},~~
 P=\prod_\rv P_\rv.
\end{equation}

% [ 1st-order ]--------------------------------------
The first-order effective Hamiltonian is derived by projecting the inter-trimer interactions
onto the degenerate manifold $V_0 = \bigotimes_\rv \mathrm{Span}(\{\ket{\dt_{\pm\frac12}}_\rv\})$.
Using Eq.~\eqref{eq:dt_trimer}, the projection of the spin operators gives the followings to first order in $\Dv$:
\begin{equation}\label{eq:S_T}
\begin{split}
 & P_\rv\Sv_{\rv,1} P_\rv=P_\rv \Sv_{\rv,3} P_\rv=\frac23 \Tv_\rv + \frac{2}{9J} \Dv_{X_\rv}\times\Tv_\rv,\\
 & P_\rv\Sv_{\rv,2} P_\rv= -\frac13 \Tv_\rv - \frac{4}{9J} \Dv_{X_\rv} \times\Tv_\rv,
\end{split}
\end{equation}
where $X_\rv=A,B$ indicates the set of trimers which $\rv$ belongs to.
Using these relations, the projection of the inter-trimer interactions gives
\begin{equation*}
\begin{split}
 PH_{\rv,\rv+\uv} P &= \Jcal_1^{\rm (1st)} \Tv_\rv\cdot\Tv_{\rv+\uv}+ \epsilon_{X_\rv} \Dcalv_1\cdot(\Tv_\rv\times\Tv_{\rv+\uv}),\\
 PH_{\rv,\rv+\vv} P &= \Jcal_1^{\rm (1st)}  \Tv_\rv\cdot\Tv_{\rv+\vv}- \epsilon_{X_\rv} \Dcalvb_1\cdot(\Tv_\rv\times\Tv_{\rv+\vv}),\\
 PH_{\rv,\rv+\bv} P &= \Jcal_2^{\rm (1st)}  \Tv_\rv\cdot\Tv_{\rv+\bv},
\end{split}
\end{equation*}
where $\epsilon_A=-\epsilon_B=1$ and
\begin{align}
 &\Jcal_1^{\rm (1st)} =\frac29 (2J_1-J'),~\Jcal_2^{\rm (1st)} =\frac89 J_2,\label{eq:Jeff1}\\
 &\Dcalv_1=\frac49 \Dv_1-\frac29 \Dv'+\frac{2}{27J} \left[ 2J_1(\Dv-\Dvb) -J'(\Dv-4\Dvb)\right].\label{eq:Deff1}
\end{align}
Notably, the effects of the three DM interactions $\Dv$, $\Dv_1$, and $\Dv'$ are now combined into the single effective DM interaction $\Dcalv_1$.
Furthermore, $\Dv_2$ has no contribution to first order
(as the contributions from the two DM interactions $\pm \Dv_{2X}$ between two trimers cancel out).
%\blue{(see also Appendix \ref{app:DM} for a symmetry argument for the vanishing of the effective DM interaction on the $\Jcal_2$ bond)}.
By using Eq.~\eqref{eq:DvDv1} and setting $\Dv'=\bm{0}$ for simplicity, we obtain the estimate
\begin{equation}\label{eq:Dcalv1}
 \Dcalv_1=(\Dcal_1^x,\Dcal_1^y,\Dcal_1^z)=(-0.5,0.8,4.8)~\mathrm{K}.
\end{equation}

%############################
\begin{table}
\caption{\label{table:estimates}
Nonzero coupling constants $\Jcal_{\Delta\rv,X}$ in the effective Hamiltonian \eqref{eq:Heff2}.
The vectors $\uv$ and $\vv$ are defined in Fig.~\ref{fig:volbeffDM}.
Only for $\Jcal_3$ and $\Jcal_3'$, the relative vector $\Delta\rv$ depends on the sublattice $X$,
as indicated in the corresponding rows.
The first- and second-order perturbative estimates (third and fourth columns)
are calculated from Eqs.~\eqref{eq:Jeff1} and \eqref{eq:Jeff2}, respectively, for the parameter set~\eqref{eq:J_volb} obtained by DFT+$U$.
}
\begin{ruledtabular}
\begin{tabular}{ccrrrr}
%$\Jcal_{\Delta\rv,X}$
& relative vectors $\Delta\rv$ & 1st-order & 2nd-order  \\
& [$(\Delta\rv,X)$ for ${\cal J}_3$, ${\cal J}_3^{\prime}$] &  &   \\
  \hline
${\cal J}_1$ & $\uv,\vv$ & $-44.8$ K & $-34.9$ K \\
${\cal J}_2$ & $\uv+\vv$ & 44.8 K & 36.5 K \\
${\cal J}_2^{\prime}$ & $-\uv+\vv$ &  & 6.8 K \\
${\cal J}_3$ & $(2\uv,A),~(2\vv,B)$ & & 4.6 K \\
${\cal J}_3^{\prime}$ & $(2\vv,A),~(2\uv,B)$ & & 1.7 K  \\
${\cal J}_4$ & $2\uv+\vv,\uv+2\uv$ &  & 1.7 K \\
${\cal J}_5$ & $2(\uv+\vv)$ &  & $-1.3$ K
% \hline
% $h_\mathrm{c1}^{(1)}~(H_\mathrm{c1}^{(1)})$ && 22.4 K (15.5 T) & 35.0 K (24.2 T) &  21.0 K (14.5 T) & 37.8 K (26.2 T)\\
% $Q/(2\pi)$ && 0.333 & 0.370 & 0.333$^*$ & 0.382\\
% $h_\mathrm{c1}^{(2)}~(H_\mathrm{c1}^{(2)})$ && 22.4 K (15.5 T) & 35.0 K (24.2 T) & 28.5 K (19.7 T) & 38.1 K (26.3 T)
\end{tabular}
\end{ruledtabular}
\end{table}
%############################

% [ 2nd-order ]--------------------------------------
The second-order contributions of the exchange couplings to the effective Hamiltonian have been calculated in Ref.~\cite{Janson16},
and found to have the magnitudes of several Kelvin, which are comparable to $|\Dcalv_1|$ above.
Taking into account these contributions also, we obtain the full effective Hamiltonian
\begin{equation}\label{eq:Heff2}
\begin{split}
 H^\mathrm{eff}=&\sum_\rv \sum_{\Delta\rv} {\cal J}_{\Delta\rv,X_\rv}\Tv_\rv \cdot \Tv_{\rv+\Delta\rv} -\hv\cdot \sum_{\rv} \Tv_\rv\\
 &+\sum_\rv \epsilon_{X_\rv} \left[ \Dcalv_1\cdot\left(\Tv_\rv\times\Tv_{\rv+\uv}\right)-\Dcalvb_1\cdot\left(\Tv_\rv\times\Tv_{\rv+\vv}\right) \right].
\end{split}
\end{equation}
The coupling constants ${\cal J}_{\Delta\rv,X}$ show seven nonzero different values as listed in Table~\ref{table:estimates},
and their second-order expressions are given by
\begin{align}\label{eq:Jeff2}
{\cal J}_1^{\rm (2nd)} =& \frac{2}{9}(2 J_1-J^\prime)
+\frac{211 {J_1}^2 + 48 J_1 J^\prime - 118 {J^\prime}^2}{1620 J} \notag\\
&+ \frac{8 J_2 (-4 J_1 + 5 J^\prime)}{243 J}, \notag\\
{\cal J}_2^{\rm (2nd)} =& \frac{8}{9} J_2 - \frac{{J_2}^2}{81 J}
-\frac{2 (-2 J_1 + J^\prime)(-13 J_1 + 8 J^\prime)}{243 J}, \notag\\
{\cal J}_2^{\prime {\rm (2nd)}} =& \frac{2(-2 J_1 + J^\prime) (-5 J_1 - 8 J^\prime)}{243 J},\notag\\
{\cal J}_3^{\rm (2nd)} =& \frac{2 (-5 J_1 + 4 J^\prime) (-J_1 - 4 J^\prime)}{243 J},\notag\\
{\cal J}_3^{\prime {\rm (2nd)}} =& \frac{5 (-2 J_1 + J^\prime)^2}{486 J},\notag\\
{\cal J}_4^{\rm (2nd)} =& \frac{8 J_2 (-4 J_1 + 5 J^\prime)}{243 J},\notag\\
{\cal J}_5^{\rm (2nd)} =& -\frac{32 {J_2}^2}{243 J}.
\end{align}
Most of these interactions do not depend on the sublattice $X$,
and have the translational invariance of the anisotropic triangular lattice.
Only the interactions $\Jcal_3$, $\Jcal'_3$, and $\Dcalv_1$ depend on the sublattice $X$, and double the unit cell of the effective model;
the primitive vectors of the system then change to $\pm\uv+\vv$.
By using Eqs.~\eqref{eq:Jeff1} and \eqref{eq:Jeff2} for the parameter set \eqref{eq:J_volb},
the effective coupling constants are estimated in Table~\ref{table:estimates}.

% [ Modification of Jcal_3 ]--------------------------------------
%As we discuss in Sec.~\ref{sec:spwv},  the parameter set in Table~\ref{table:estimates} and Eq.~\eqref{eq:Dcalv1}
%does not provide a full quantitative description of volborthite.
%In the search for a suitable parameter set for volborthite, we have found it useful to allow a slight modification of the $\Jcal_3$ coupling from the second-order expression as follows:
%\begin{equation}
% \Jcal_3=\Jcal_3^{\rm (2nd)} + \delta\Jcal_3.
%\end{equation}

Although the effective coupling constants other than $\Jcal_1$ and $\Jcal_2$ in Table~\ref{table:estimates} have comparatively small magnitudes,
they can significantly influence the low-temperature properties of physical quantities and the phase diagram
as we discuss in Sections~\ref{sec:spwv} and \ref{sec:fieldtheory}.
To examine such sensitivity to the small effective coupling constants,
we allow a slight modification of the $\Jcal_3$ coupling from the second-order expression as
\begin{equation}\label{eq:dJcal3}
 \Jcal_3=\Jcal_3^{\rm (2nd)} + \delta\Jcal_3.
\end{equation}
This modification is also useful in finding a parameter set consistent with the experimental data of the magnetization and the thermal Hall conductivity,
as we discuss in Sec.~\ref{sec:constraints_J}.
Microscopically, $\delta\Jcal_3$ can arise from the second-neighbor interaction $J_2'$ along the direction of the $J'$ bond
in the original model---this interaction can be written down as
\begin{equation}
 J_2' \left[ \sum_{\rv\in A} \Sv_{\rv,3}\cdot\Sv_{\rv+2\uv,1} + \sum_{\rv\in B} \Sv_{\rv,3}\cdot\Sv_{\rv+2\vv,1} \right],
\end{equation}
and the first-order perturbation theory gives $\delta\Jcal_3=(4/9) J_2'$.

%%%%%%%%%%%%%%%%%%%%%%%%%%%%%%%%%%%%%%%%%%%%%%%%%
\section{Spin wave analysis from the $\frac13$-plateau state} \label{sec:spwv}
%%%%%%%%%%%%%%%%%%%%%%%%%%%%%%%%%%%%%%%%%%%%%%%%%

In this section, we analyze the effective pseudospin-$\frac12$ Hamiltonian \eqref{eq:Heff2} at high fields by using the spin wave theory.
We mainly consider the case in which the field is applied in the $-z$ direction [see Eq.~\eqref{eq:H_z}].
In experiments for single crystals, a broad magnetization plateau at $\frac13$ of the saturation
has been found to appear for $H>H_\cone\simeq 27.5$~T \cite{Ishikawa15,Nakamura18}.
In the effective model, this corresponds to the fully polarized state of pseudospins.
Near but below the $\frac13$-plateau phase, the system can be viewed as a finite-density gas of magnons.
We consider the temperature regime where these magnons are not condensed or forming bimagnon bound states.
We show that magnon Bloch states acquire a nonzero Berry curvature
through a combination of the effective Heisenberg and DM interactions, giving rise to
a thermal Hall effect \cite{Onose10,Katsura10, Matsumoto11_PRL, Matsumoto11_PRB, Murakami17}.
We calculate the field- and temperature-dependences of the magnetization and the thermal Hall conductivity.
Comparison with experimental data \cite{Ishikawa15,Watanabe16} indicates the necessity to modify
the effective coupling constants significantly from the DFT-based estimates (Table \ref{table:estimates} and Eq.\ \eqref{eq:Dcalv1}),
and we discuss some constraints imposed on these constants.
We also discuss the cases of other field directions briefly.
We note that our analysis of the thermal Hall effect based on the effective model
sharply contrasts with those based on isotropic kagome models \cite{Owerre17_EPL,Doki18,GaoYH19}.

%We note that the spin wave analysis from the $\frac13$-plateau state has been done for the coupled frustrated chain model in Ref.\ \cite{Parker17}.

%================================================
\subsection{Spin wave Hamiltonian} \label{sec:spwvH}
%================================================

We consider the effective pseudospin-$\frac12$ Hamiltonian \eqref{eq:Heff2} with $\hv=(0,0,h)$~($h>0$).
The plateau state at high fields $h$ corresponds to the fully polarized state of pseudospins
$\bigotimes_\rv \ket{\dt_{+\frac12}}_\rv$, which we view as the magnon vacuum in the following.
Then, $T^\pm_\rv=T^x_\rv\pm i T^y_\rv$ play the role of magnon annihilation and creation operators,
and the magnon occupation number at the site $\rv$ is given by $n_\rv=T_\rv^- T_\rv^+=\frac12-T_\rv^z$.
In the following, we replace $T^+_\rv$ by the bosonic annihilation operator $a_\rv$ satisfying the commutation relation $[a_\rv,a_{\rv'}^\dagger]=\delta_{\rv\rv'}$,
and introduce an infinite on-site repulsion $U_0\to\infty$ to impose the hard-core constraint.
Furthermore, concerning the DM interactions, we only take into account the $z$ component of the DM vectors to simplify the analysis.
This treatment is justified below 
the lower edge of the $\frac13$-plateau (i.e., the saturation field of the pseudospin-$\frac12$ model)
and at low temperatures
as the $x$ and $y$ components of the DM vectors vanish in the low-energy effective field theory as we see in Sec.~\ref{sec:fieldtheory}.

Our effective Hamiltonian \eqref{eq:Heff2} can thus be rewritten in terms of the bosonic operators as
\begin{equation}\label{eq:Heff_Hmag}
 H^\mathrm{eff}=N_\mathrm{t} \left( \frac{\Jcal}4 -\frac{h}{2} \right) + h \sum_\rv n_\rv + H_\kinetic + H_\interact,
\end{equation}
where $\Ntri$ is the number of trimers in the system and
\begin{align}
 \Jcal &= \frac1{\Ntri} \sum_\rv \sum_{\Delta\rv} \Jcal_{\Delta\rv,X_\rv} \notag\\
 &=2\Jcal_1+\Jcal_2+\Jcal_2'+\Jcal_3+\Jcal_3'+2 \Jcal_4+\Jcal_5,\\
 H_\kinetic &= -\Jcal \sum_\rv n_\rv
 + \sum_\rv \sum_{\Delta\rv} \frac{ {\cal J}_{\Delta\rv,X_\rv} }{2}
 \left( a^\dagger_{\rv+\Delta\rv} a_\rv + \mathrm{h.c.} \right) \notag\\
 &~~~+ \Dcal_1^z\sum_\rv \epsilon_{X_\rv} \sum_{\Delta\rv=\uv,\vv} \frac{i}{2} \left( a^\dagger_{\rv+\Delta\rv} a_\rv-a^\dagger_\rv a_{\rv+\Delta\rv} \right),\label{eq:Hmag_kin}\\
 H_\interact & = \frac{U_0}2 \sum_\rv n_\rv(n_\rv-1) + \sum_\rv \sum_{\Delta\rv} {\cal J}_{\Delta\rv,X_\rv} n_\rv n_{\rv+\Delta\rv}. \label{eq:Hmag_int}
\end{align}
The first term in Eq.~\eqref{eq:Heff_Hmag} is the energy of the magnon vacuum.
The second term is the Zeeman term, which plays the role of a magnon chemical potential.
The third and fourth terms, $H_\kinetic$ and $H_\interact$, are magnon kinetic and interaction energies, which do not depend on $h$.
We note that the total number of magnons, $\sum_\rv n_\rv$, is conserved in this Hamiltonian since we have neglected the $x$ and $y$ components of the DM vectors.

We comment that the $\Jcal_1$ and $\Dcal_1^z$ terms in $H_\kinetic$ can be combined into the form
\begin{equation}\label{eq:Hkin_J1p}
-\sum_\rv \sum_{\Delta\rv=\uv,\vv} \frac12 \sqrt{{\Jcal_1}^2+(\Dcal_1^z)^2} \left(e^{-i\epsilon_{X_\rv}\nu} a_{\rv+\Delta\rv}^\dagger a_\rv+\mathrm{h.c.}\right) ,
\end{equation}
where $\nu:=\arg (-\Jcal_1+ i\Dcal_1^z)$.
When $\Jcal_4=0$, the phase factor $e^{-i\epsilon_{X_\rv}\nu}$ in Eq.\ \eqref{eq:Hkin_J1p} can be removed
by performing the gauge transformation  $a'_\rv=e^{i\nu} a_\rv$ for $\rv\in B$
while keeping the other terms in the Hamiltonian unchanged.
However, such removal of phase factors from $H_\kinetic$ is not possible for $\Jcal_4\ne 0$
as the $\Jcal_4$ term acquires phase factors by the above gauge transformation 
(see Refs.\ \cite{Kaplan83, Shekhtman92} for related discussions in other contexts).
Therefore, the presence of $\Jcal_4$ is crucial for finding nontrivial effects of the complex hopping amplitude of magnons induced by $\Dcal_1^z$.

%================================================
\subsection{Magnon Bloch states}\label{sec:Bloch}
%================================================

% [ Magnon Bloch bands ]--------------------------------------
We first analyze the kinetic part $H_\kinetic$ [Eq.~\eqref{eq:Hmag_kin}] of the spin wave Hamiltonian
and determine the Bloch states of magnons. By performing the Fourier expansion
\begin{equation}
 a_\rv = \frac{1}{\sqrt{\Ntri /2}} \sum_\kv e^{i\kv\cdot\rv} a_{\kv X}~~(\rv\in X),
\end{equation}
where the sum is over the discrete wave vectors in the first Brillouin zone,
$H_\kinetic$ can be rewritten as
\begin{equation}
 H_\mathrm{kin} = \sum_\kv \left( a_{\kv A}^\dagger, a_{\kv B}^\dagger \right) M(\kv)
 \begin{pmatrix} a_{\kv A} \\ a_{\kv B} \end{pmatrix} .
\end{equation}
Here, the $2\times 2$ matrix $M(\kv)$ is given by
\begin{equation}
 M(\kv) = E (\kv) I + \bm{J} (\kv) \cdot \bm{\sigma},
\end{equation}
where $I$ is the identity matrix, $\bm{\sigma}=(\sigma^x,\sigma^y,\sigma^z)$ are the Pauli matrices, and
\begin{equation}\label{eq:epsilon_J_k}
\begin{split}
 E (\kv) &= -\Jcal
 +\Jcal_2\cos (k_u+k_v) + \Jcal_2' \cos(k_u-k_v) \\
 &~~~+ \frac12 (\Jcal_3+\Jcal_3') \left[ \cos(2k_u)+\cos(2k_v) \right] \\
 &~~~+ \Jcal_5\cos(2k_u+2k_v), \\
 J^x(\kv)&= \Jcal_1 \left( \cos k_u + \cos k_v \right) \\
 &~~~ + \Jcal_4 \left[ \cos(2k_u+k_v) + \cos(k_u+2k_v) \right], \\
 J^y(\kv)&= \Dcal_1^z \left( \cos k_u + \cos k_v \right),\\
 J^z(\kv)&= \frac12 (\Jcal_3-\Jcal_3') \left[ \cos(2k_u)-\cos(2k_v) \right]
\end{split}
\end{equation}
with $k_u=\kv\cdot\uv$ and $k_v=\kv\cdot\vv$.
The two Bloch energy bands are calculated as
\begin{equation}\label{eq:epsilon_pm_k}
 E_\pm (\kv) = E (\kv) \pm |\bm{J}(\kv)|.
\end{equation}
It is easy to show that the two bands touch along the entire boundary of the Brillouin zone ($|k_x|=\pi/b$ or $|k_y|=\pi/a$),
which can be interpreted as the Kramers degeneracy due to certain antiunitary symmetries
as explained in Appendix \ref{app:band_touch}.
By parameterizing the vector $\Jv (\kv)$ in terms of the polar coordinates as
\begin{equation}\label{eq:J_k}
 \Jv (\kv) = |\Jv(\kv)| \left( \sin\theta(\kv)\cos\phi(\kv), \sin\theta(\kv)\sin\phi(\kv), \cos\theta(\kv) \right),
\end{equation}
the eigenstates corresponding to Eq.~\eqref{eq:epsilon_pm_k} are given by
\begin{equation}\label{eq:psi_pm}
\begin{split}
 \ket{\psi_+(\kv)} &= \begin{pmatrix} \cos(\theta(\kv)/2) \\ e^{i\phi(\kv)} \sin(\theta(\kv)/2) \end{pmatrix},\\
 \ket{\psi_-(\kv)} &= \begin{pmatrix} -\sin(\theta(\kv)/2) \\ e^{i\phi(\kv)} \cos(\theta(\kv)/2) \end{pmatrix}.
\end{split}
\end{equation}

%############################
\begin{figure}
\begin{center}
\includegraphics[width=0.48\textwidth]{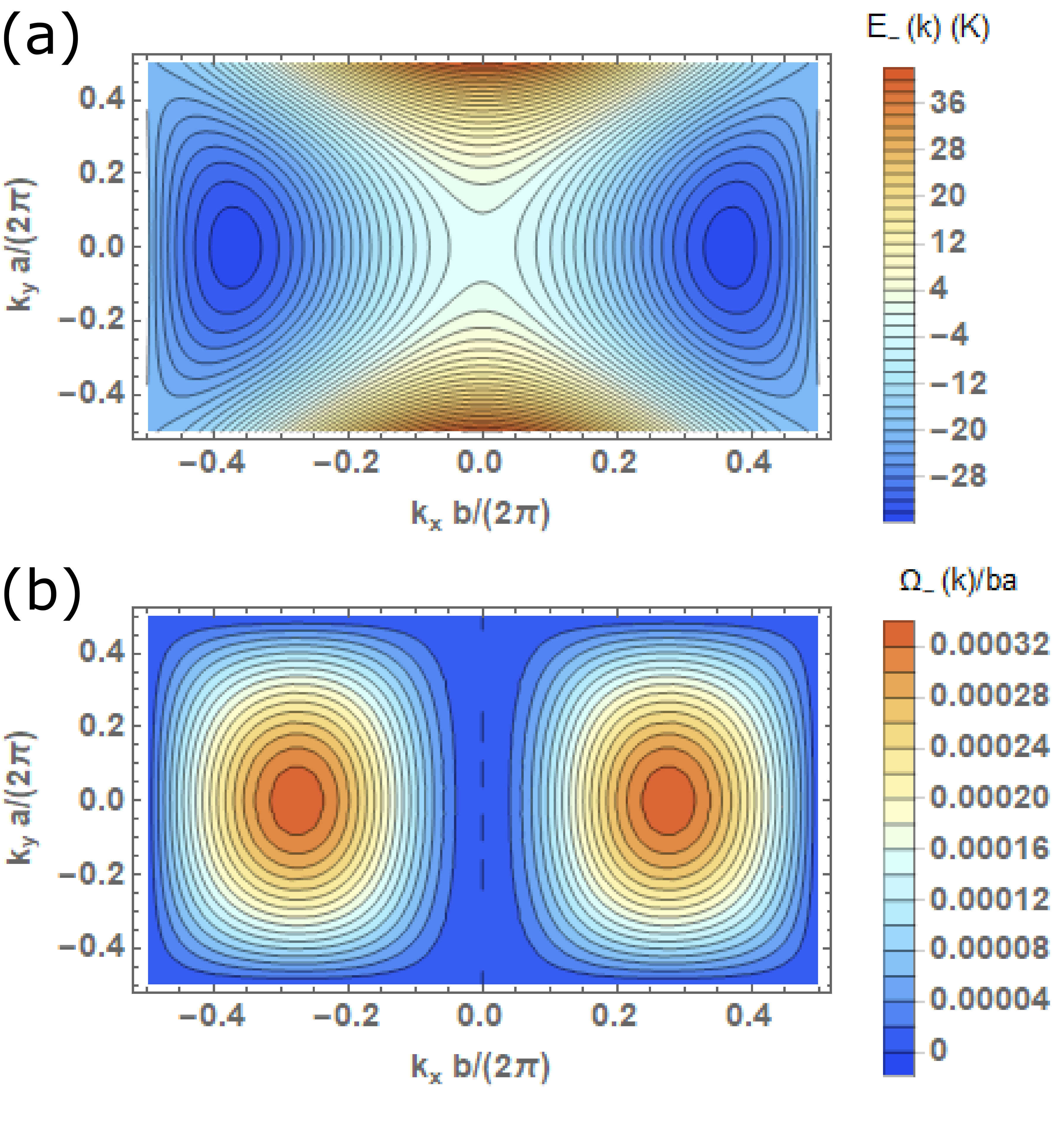}
\end{center}
\caption{\label{fig:B_dz}
(Color online)
(a) The energy $E_-(\kv)$ [Eq.~\eqref{eq:epsilon_pm_k}] and
(b) the dimensionless Berry curvature $\Omega_- (\kv)/(ba)$ [Eqs.~\eqref{eq:Omega_psi} and \eqref{eq:Omega_theta_phi}]
for the lower magnon band over the first Brillouin zone.
These are calculated for the second-order model (see Table \ref{table:estimates})
with the value of $\Dcal_1^z$ in Eq.\ \eqref{eq:Dcalv1}.
}
\end{figure}
%############################

% [ Magnon BEC ]--------------------------------------
Taking account of the second term (Zeeman energy) in Eq.\ \eqref{eq:Heff_Hmag} also,
we find that the single-magnon Bloch state has an excitation energy $E_\pm(\kv)+h$ measured from the vacuum (the $\frac13$-plateau state).
Introducing $E_0:= \min_\kv E_-(\kv)$, we find that the magnon excitation energy is gapped over the entire Brillouin zone for $h>h_\cone:= -E_0$.
At $h=h_\cone$, the bottom of the lower magnon band touches zero, leading to a Bose-Einstein condensation of magnons \cite{Zapf14,Nikuni00} at zero temperature.
For $h<h_\cone$, repulsion between magnons stabilizes a finite-density condensate of magnons.
Since the interaction part $H_\interact$ [Eq.~\eqref{eq:Hmag_int}] of the spin wave Hamiltonian has no contribution in the single-magnon problem,
$h_\cone$ gives the exact single-magnon condensation point.

In systems with competing ferromagnetic and antiferromagnetic interactions, however,
multi-magnon bound states are formed under certain conditions
and can condense before the single-magnon states do with lowering the field $h$.
A condensation of two-magnon bound states (bimagnons) gives rise to a bond nematic order \cite{Momoi05,Shannon06,Kecke07,Hikihara08,Sudan09,Balents16}.
In Ref.~\cite{Janson16}, it has been found that while the first- and second-order models in Table \ref{table:estimates} show a conventional single-magnon condensation,
slightly modified models with reduced $\Jcal_4$ show a condensation of bimagnons.
Here we do not address the nature of the low-temperature phase below the $\frac13$-plateau phase in further detail in this section.
We are instead interested in the thermodynamic behavior of magnons at finite temperatures
well above the regime where a single- or two-magnon condensation occurs or magnon bound states are formed.

% [ Lower band ]--------------------------------------
The lower energy band $E_-(\kv)$ is plotted in Fig.~\ref{fig:B_dz}(a)
for the second-order model (see Table \ref{table:estimates}) with the value of $\Dcal_1^z$ in Eq.\ \eqref{eq:Dcalv1}.
It shows the minimum value $E_0=-35.2$~K at $\kv =\pm (Q/b,0)$ with $Q/(2\pi)=0.369$.
The single-magnon condensation point is therefore given by $h_\cone=35.2~\mathrm{K}=(g\mu_B/k_B)\times 24.4$ T,
which reasonably agrees with the low-field end $H_\cone\simeq 27.5$ T of the $\frac13$-plateau observed experimentally \cite{Ishikawa15,YoshidaM17,Kohama19}.

For a later purpose, we further consider the expansion of $E_-(\kv)$ around the minima $\kv=(\pm Q/b,0)$:
\begin{equation}\label{eq:em_expand}
 E_-(\kv)\approx E_0+\frac{C_x}2 \left(k_xb\mp Q\right)^2 + \frac{C_y}2  \left( k_y a \right)^2,
\end{equation}
where the coefficients are given by $(C_x,C_y)=(42.8, 5.12)$~K for the case of Fig.~\ref{fig:B_dz}(a).
This leads to a constant density of states $G=1/(\pi \sqrt{C_xC_y})=0.0215$~K$^{-1}$ in units of $\Ntri/2$ at low energies.

%$h_\cone=35.2$ K or $H_\cone=24.4$ T,
%hc1=35.2387 K, Hc1=24.3891 T

%================================================
\subsection{Magnetization}\label{sec:spwv_magh}
%================================================

% [ Effective chemical potential for magnons ]--------------------------------------
To study a finite-density gas of magnons,
it is important to treat the magnon interaction part $H_\interact$ in Eq.~\eqref{eq:Hmag_int} properly.
To this end, we perform the mean-field decoupling of this part,
which results in the effective chemical potential $\mu=-(h-h_\cone)-2Un$ for magnons \cite{Nikuni00}.
Here, $n$ is the magnon number per trimer, and $U$ is the effective interaction parameter
which encompasses the effects of all the interaction terms in Eq.~\eqref{eq:Hmag_int}.
Furthermore, $\mu$ is measured relative to $E_0=-h_\cone$ so that the condensation occurs at $\mu=0$.
Since there is a subtlety in the mean-field treatment of the infinite on-site interaction $U_0$,
it is challenging to determine $U$ through a microscopic calculation.
We instead determine $U$ later in such a way that consistency with the experimental magnetization data \cite{Ishikawa15} is achieved.
The magnon density $n$ is obtained as a function of $\mu$ and $T$ as
\begin{equation}\label{eq:n_T_mu}
 n(\mu,T) = \frac{1}{\Ntri} \sum_\rv \langle n_\rv\rangle = \frac12 \sum_{\alpha=\pm} \int_\BZ \frac{d^2\kv ba}{(2\pi)^2} \rho_\alpha(\kv),
\end{equation}
where $\rho_\alpha (\kv)=\left[ e^{(E_\alpha(\kv)-E_0-\mu)/T}-1 \right]^{-1}$ is the Bose distribution function and the integration is over the Brillouin zone.
A finite magnon density $n>0$ leads to a reduction in the magnetization from the $\frac13$-plateau.

% [ How to draw magnetization curve ]--------------------------------------
In the experimental magnetization data $M(H,T)$ for $T=1.4$ K,
a magnetization plateau with $M_\mathrm{p}=0.38\mu_B$ (per Cu$^{2+}$ ion) has been found
after subtracting the van Vleck contribution $M_\mathrm{VV}=0.000146~(\Tesla^{-1})\times \mu_B H$ \cite{Ishikawa15}.
The deviation from the plateau, $\Delta M := M-M_\mathrm{VV}-M_\mathrm{p}$,
should be related with the magnon density $n$ as
\begin{equation}\label{eq:M_n}
 \Delta M = -g\mu_B n(\mu,T)/3,
\end{equation}
where the division by $3$ comes from the fact that $n$ is defined per trimer.
The magnetic field $H$ is related with the chemical potential $\mu$ and the magnon density $n$ as
\begin{equation}\label{eq:H_mu_n}
 H-H_\cone = \frac{k_B}{g\mu_B} (h-h_\cone) = -\frac{k_B}{g\mu_B} [\mu+2Un(\mu,T)].
\end{equation}
For given $T$ and $U$, we can thus obtain the theoretical magnetization curve around $H=H_\cone$
by calculating $n(\mu,T)$ in Eq.~\eqref{eq:n_T_mu} as a function of $\mu<0$ and plotting the relation between Eqs.\ \eqref{eq:M_n} and \eqref{eq:H_mu_n}.
Although the values of $H_\cone$ are slightly different between theory and experiment (as noted in Sec.~\ref{sec:Bloch}),
the magnetization curves as a function of $H-H_\cone$ can be compared in a quantitative manner.

%############################
\begin{figure}
\begin{center}
\includegraphics[width=0.5\textwidth]{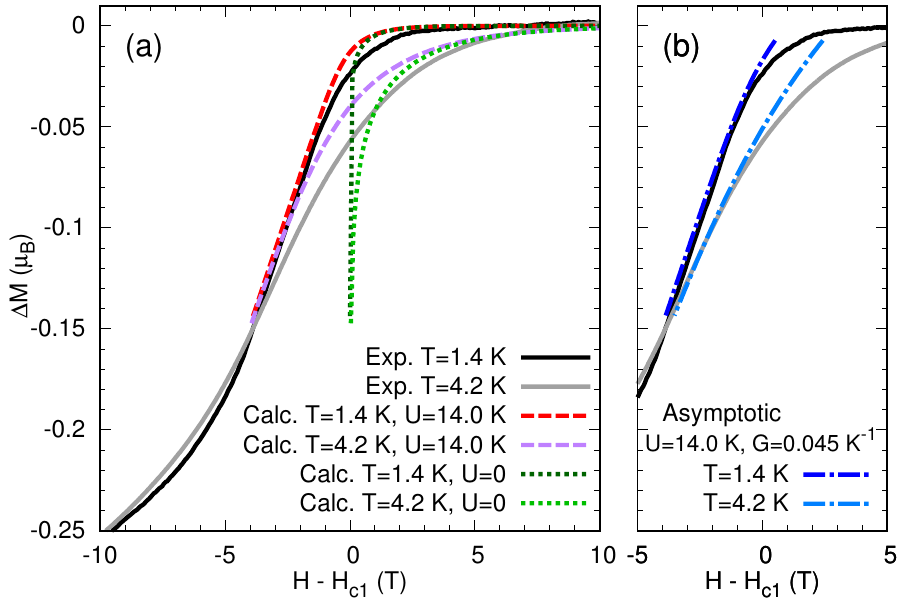}
\end{center}
\caption{\label{fig:magh}
(Color online)
(a) Comparison of magnetization curves between experiment \cite{Ishikawa15} and our theoretical calculations for $T=1.4$ K and $4.2$ K.
The magnetization is measured relative to the $\frac13$-plateau
while the field $H$ is relative to the low-field end $H_\cone$ of the plateau.
The theoretical calculations are performed by using Eqs.\ \eqref{eq:M_n} and \eqref{eq:H_mu_n} for the same model as in Fig.~\ref{fig:B_dz}.
The effective interaction parameter $U=14.0$ K is obtained so as to reproduce
the slope of the $T=1.4$ K experimental curve slightly below the plateau [see Appendix \ref{app:EstU}].
When $U$ is set to zero, the magnon density $n\propto -\Delta M$ diverges as $H\searrow H_\cone$.
(b) Fit of the experimental data with the asymptotic form \eqref{eq:h_n_asymp}.
This yields an estimate $G=0.045$~K$^{-1}$ of the low-energy density of states of magnons.
}
\end{figure}
%############################

% [ Magnetization curves: experiment vs. theory ]--------------------------------------
In Fig.\ \ref{fig:magh}(a), we compare the theoretical and experimental magnetization curves for $T=1.4$ K and $4.2$ K.
In the absence of an interaction ($U=0$), the magnon density $n\propto -\Delta M$ diverges with lowering the field $H$ to $H_\cone$.
A finite magnon density $n\propto -\Delta M$ for $H<H_\cone$ in the experimental data can therefore be interpreted as a result of magnon repulsion.
We determine the effective interaction parameter $U$ in such a way that
the slope of the $T=1.4$ K experimental magnetization curve slightly below the $\frac13$-plateau is reproduced
(see Appendix \ref{app:EstU}).
With $U=14.0$ K, we indeed find good agreement between theoretical and experimental curves for $-4~\Tesla<H-H_\cone<-2~\Tesla$ and $T=1.4$ K.
For $-2~\Tesla<H-H_\cone$ or $T=4.2$ K, however, the magnon density $n\propto -\Delta M$ tends to be larger in experiment than in theory.
This indicates that the density of states of magnons from the $\frac13$-plateau state in volborthite
is larger than that expected in the effective model with the present parameter values.
This simply implies that volborthite is more frustrated than our present model.

We can indeed estimate the density of states of magnons by fitting the experimental data using the asymptotic form
\begin{equation}\label{eq:h_n_asymp}
 h-h_\cone = -2Un +T \exp\left( - \frac{2n}{GT} \right) ~~ \left( n\gtrsim GT/2 \right),
\end{equation}
which is derived in Appendix \ref{app:EstU}.
Here, $G$ is the density of states of magnons in units of $\Ntri/2$ in the low-energy limit.
The best fit as shown in Fig.\ \ref{fig:magh}(b) gives $G=0.045$ K$^{-1}$,
which is roughly twice as large as the value $G=0.0215$ K$^{-1}$ for the model used in Fig.\ \ref{fig:magh}(a).
This indicates the necessity to modify the effective coupling constants $\{\Jcal_{\Delta \rv,X}\}$
to have consistency with experiment, as we discuss in more detail in Sec.\ \ref{sec:constraints_J}.

%================================================
\subsection{Thermal Hall conductivity}
%================================================

\begin{figure}
\begin{center}
\includegraphics[width=0.5\textwidth]{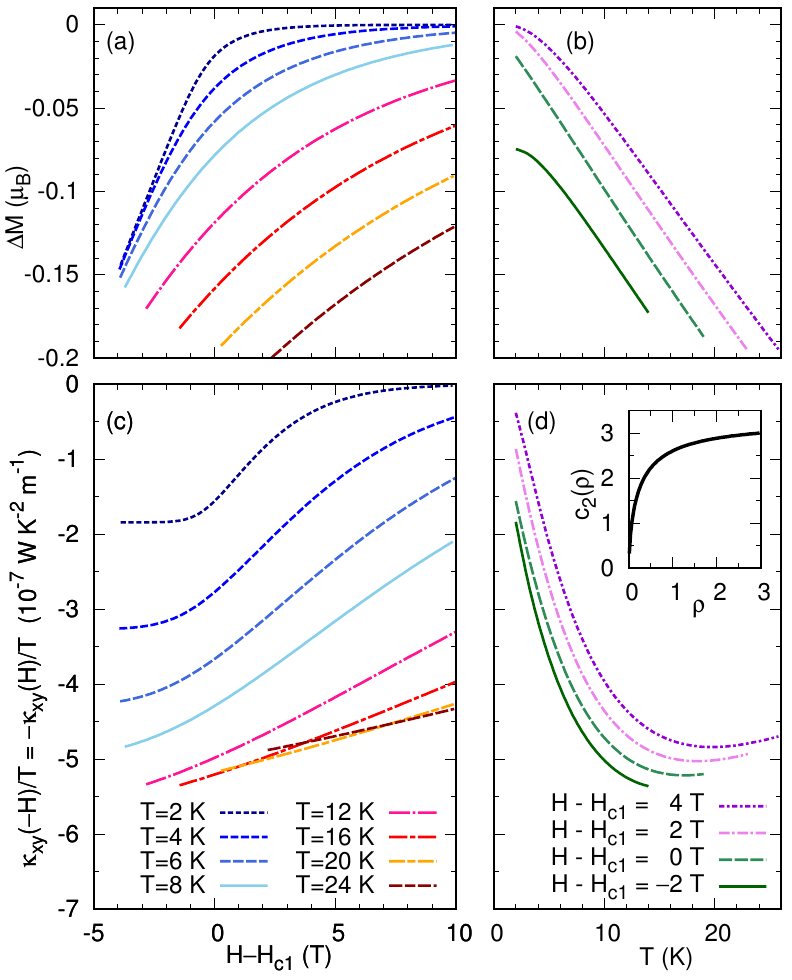}
\end{center}
\caption{\label{fig:kappaoT}
(Color online)
(a,b) The magnetization (measured from the $\frac13$-plateau as in Fig.~\ref{fig:magh})
and (c,d) the thermal Hall conductivity $\kappa_{xy}(-H)/T=-\kappa_{xy}(H)/T$
as functions of $H-H_\cone$ and $T$ for the same model as in Fig.~\ref{fig:B_dz}.
The inset of (d) shows the function $c_2(\rho)$ in Eq.~\eqref{eq:ctwo}.
}
\end{figure}

% [ Berry phase formula of the thermal Hall conductivity ]--------------------------------------
The thermal Hall conductivity in the clean limit is given by \cite{Matsumoto11_PRL,Matsumoto11_PRB,Murakami17}
\begin{equation}\label{eq:kappaxy}
 \kappa^{xy} = - \frac{k_B^2 T}{\hbar V}\cdot 2N_c \sum_{\alpha, \kv} c_2(\rho_\alpha(\kv)) \Omega_\alpha (\kv),
\end{equation}
where $2N_c$ is the number of layers in the system, and $V=(\Ntri/2)ab\cdot N_c c\sin\beta$ is the volume of the system
(we remind the reader that two inequivalent layers are alternately stacked in the $c$ direction in volborthite).
Here, we assume that inter-layer interactions can be neglected and the contributions from different layers can simply be summed up, hence the multiplication by $2N_c$ in Eq.\ \eqref{eq:kappaxy}.
The function $c_2(\rho)$ is given by
\begin{equation}\label{eq:ctwo}
 c_2(\rho)
 = (1+\rho) \left( \ln \frac{1+\rho}{\rho} \right)^2
 - \left(\ln \rho \right)^2 -2\mathrm{Li}_2(-\rho),
\end{equation}
where $\mathrm{Li}_m(z)=\sum_{n=1}^\infty z^n/n^m$ is the polylogarithm function;
see the inset of Fig.~\ref{fig:kappaoT}(d) for a plot of $c_2(\rho)$.
The Berry curvature $\Omega_\alpha(\kv)~(\alpha=\pm)$ is defined as
\begin{equation}\label{eq:Omega_psi}
 \Omega_\alpha (\kv) = i \sum_{i,j} \epsilon_{ij} \bracket{\partial_i\psi_\alpha}{\partial_j\psi_\alpha},
\end{equation}
where $\partial_i=\frac{\partial}{\partial k_i}~(i=x,y)$ and
$\epsilon_{ij}$ is an antisymmetric tensor with $\epsilon_{xy}=-\epsilon_{yx}=1$.

% [ Berry curvature ]--------------------------------------
The dimensionless Berry curvature $\Omega_- (\kv)/(ba)$ for the lower band is plotted in Fig.~\ref{fig:B_dz}(b).
It shows the maximal value $\Omega_- (\kv)/(ba)\simeq 3\times 10^{-4}$ near $\kv=\pm \kv_*:=\pm(\pi/(2b),0)$,
and vanishes at the Brillouin zone boundary.
At $\kv=\kv_*$, the following simple expression of the Berry curvature is available (see Appendix \ref{app:Berry} for the derivation):
\begin{equation}\label{eq:Omega_kstar}
 \frac{\Omega_\pm (\kv_*)}{ba} = \mp \frac{(\Jcal_3-\Jcal_3')\Jcal_4\Dcal_1^z}{\sqrt{2} \left[ (\Jcal_1-\Jcal_4)^2+(\Dcal_1^z)^2 \right]^{3/2}}.
\end{equation}
This expression suggests that the sign and the magnitude of the Berry curvature are controlled not only by the DM interaction $\Dcal_1^z$
but also by the long-range effective couplings $\Jcal_3-\Jcal_3'$ and $\Jcal_4$ which arise from the second-order strong-coupling expansion.
Furthermore, the magnitude of the Berry curvature in Eq.~\eqref{eq:Omega_kstar} depends significantly on the nearest-neighbor effective coupling $\Jcal_1$.

The importance of the effective coupling constants discussed above for the magnitude of the Berry curvature can be understood as follows.
In order to have nonzero $\Omega_\pm(\kv)$, the vector $\Jv(\kv)$ in Eq.~\eqref{eq:epsilon_J_k} must form a solid angle as we vary $\kv$ in different directions.
When $\Dcal_1^z=0$ or $\Jcal_3-\Jcal_3'=0$, one component of $\Jv(\kv)$ vanishes and $\Jv(\kv)$ is thus constrained to a 2D plane.
When $\Jcal_4=0$, we have $\Jcal^x(\kv)\propto \Jcal^y(\kv)$, and $\Jv(\kv)$ is again constrained to a 2D plane.
Therefore, $\Dcal_1^z$, $\Jcal_3-\Jcal_3'$, and $\Jcal_4$ must all be nonzero to have nonzero $\Omega_\pm(\kv)$ and $\kappa^{xy}$.
The importance of these coupling constants can also be understood from the following viewpoints:
(i) When $\Jcal_3-\Jcal_3'=\Dcal_1^z=0$, the effective model acquires the invariance under the translations by $\uv$ and $\vv$,
and no longer satisfies minimal requirement of a two-sublattice structure for a finite Berry curvature;
(ii) When $\Jcal_4=0$, the complex hopping amplitude of magnons due to $\Dcal_1^z$ can be transformed into a real one
by a gauge transformation as discussed in Sec.\ \ref{sec:spwvH}.
We note that the dimensionless Berry curvature $\Omega_- (\kv)/(ba)$ in Fig.~\ref{fig:B_dz}(b) takes only small values of order $10^{-4}$ over the entire Brillouin zone
because the vector $\Jv(\kv)$ changes only around the $-x$ direction owing to $-\Jcal_1\gg |\Jcal_3-\Jcal_3'|,|\Jcal_4|,|\Dcal_1^z|$ in the present model.

%Therefore, the thermal Hall conductivity has the order of $\kappa_{xy}/T\sim 10^{-7}~\mathrm{W}\cdot\mathrm{K}^{-2}\cdot\mathrm{m}^{-1}$.

%############################
%############################

% [ kappa_xy ]--------------------------------------
To evaluate $\kappa^{xy}$, it is useful to rewrite Eq.~\eqref{eq:kappaxy} as
\begin{equation}\label{eq:kappaxy_integral}
 \frac{\kappa^{xy}}{T} = - \frac{2k_B^2}{\hbar c\sin\beta} \sum_{\alpha=\pm} \int_\BZ \frac{d^2\kv}{(2\pi)^2} c_2(\rho_\alpha(\kv))  \Omega_\alpha (\kv),
\end{equation}
where the coefficient is evaluated as
\begin{equation}
 \frac{2k_B^2}{\hbar c\sin\beta} = 2.521\times 10^{-3}~\mathrm{W}\cdot\mathrm{K}^{-2}\cdot\mathrm{m}^{-1}.
\end{equation}
We note that $\kappa^{xy}/T$ is an antisymmetric function of the magnetic field $H$: $\kappa^{xy}(-H)/T=-\kappa^{xy}(H)/T$.
Since we have assumed a field $\Hv=(0,0,-H)$ in the $-z$ direction [see Eq.~\eqref{eq:H_z}],
our calculation based on Eq.~\eqref{eq:kappaxy_integral} evaluates the left-hand side of this relation.

In Fig.~\ref{fig:kappaoT}, we plot the magnetization $\Delta M$ relative to the $\frac13$-plateau
and the thermal Hall conductivity $\kappa^{xy}(-H)/T=-\kappa^{xy}(H)/T$ as functions of $H-H_\cone$ and $T$.
As we lower the field $H$ or raise the temperature $T$, the magnon density $n\propto -\Delta M$ increases.
At the same time, $|\kappa^{xy}(-H)|/T$ tends to increase although a non-monotonic dependence on $T$ starts around $T=16$~K.
In particular, $|\kappa^{xy}(-H)|/T$ shows a peak as a function of $T$ as seen in Fig.~\ref{fig:kappaoT}(d).
The decrease in $|\kappa^{xy}(-H)|/T$ at high temperatures can be understood by noting that
the upper- and lower-band contributions gradually cancel out in Eq.~\eqref{eq:kappaxy_integral}
owing to the zero sum rule for the Berry curvature $\sum_\alpha \Omega_\alpha(\kv)=0$ \cite{Xiao10} [see also Eq.\ \eqref{eq:Omega_theta_phi}]
and slower variation of the function $c_2(\rho)$ for a larger number $\rho$ of magnons [see the inset of Fig.~\ref{fig:kappaoT}(d)].

Let us now compare our results with the experimental data of Watanabe {\it et al.}\ \cite{Watanabe16} for $|H|\lesssim 15$ T.
Although the experimental field range is away from the $\frac13$-plateau,
it is expected that magnitudes and qualitative features of the thermal Hall conductivity do not change abruptly
as we vary $|H|$ at sufficiently high temperatures $T\gtrsim 2$ K where there is no phase transition.
The appearance of a peak in the $T$ dependence of $|\kappa^{xy}(-H)|/T$ in Fig.~\ref{fig:kappaoT}(d)
is qualitatively consistent with the experimental data for $H=15$ T.
However, the experimental peak value $\kappa^{xy}(-H)/T\sim 4\times 10^{-5}~\mathrm{W}\cdot\mathrm{K}^{-2}\cdot\mathrm{m}^{-1}$ is
two orders of magnitudes larger than our data in Fig.~\ref{fig:kappaoT}(c,d).
Furthermore, our data in Fig.~\ref{fig:kappaoT}(c,d) always show $\kappa^{xy}(-H)<0$
while the experimental data show $\kappa^{xy}(-H)>0$ for $T\gtrsim 4$~K.
We note that the sign and the magnitude of the thermal Hall conductivity
depend crucially on various effective coupling constants
as seen in the representative value of the Berry curvature in Eq.\ \eqref{eq:Omega_kstar}.
Specifically, the magnitude of the right-hand side of Eq.\ \eqref{eq:Omega_kstar} significantly increases
as we weaken $\Jcal_1$ from the present estimate.
We discuss this issue further in Sec.\ \ref{sec:constraints_J}.

% [ Chern number ]--------------------------------------
In passing, we note that in general, the integral of the Berry curvature over the Brillouin zone for an {\it isolated} band,
$\mathcal{N}_\alpha:= \int_\BZ\frac{d^2\kv}{2\pi} \Omega_\alpha(\kv)$,  is topologically quantized to integers \cite{Thouless82}.
In the present case, however, the two bands touch at the Brillouin zone boundary, and thus the quantization rule does not apply individually for each band.
Indeed, in Fig.~\ref{fig:B_dz}(b), $\Omega_- (\kv)/(ba)$ shows positive but small values over the Brillouin zone,
and the full integration of it gives only a tiny non-integer value $\mathcal{N}_-=7.5\times 10^{-4}$.
This contrasts with the case of SrCu$_2$(BO$_3$)$_2$,
for which a theory predicts the emergence of topologically nontrivial bands and an associated large thermal Hall conductivity
through a combination of DM interactions and a magnetic field \cite{Romhanyi15}.

%================================================
\subsection{Modification of the coupling constants} \label{sec:constraints_J}
%================================================

%############################
\begin{figure}
\begin{center}
\includegraphics[width=0.48\textwidth]{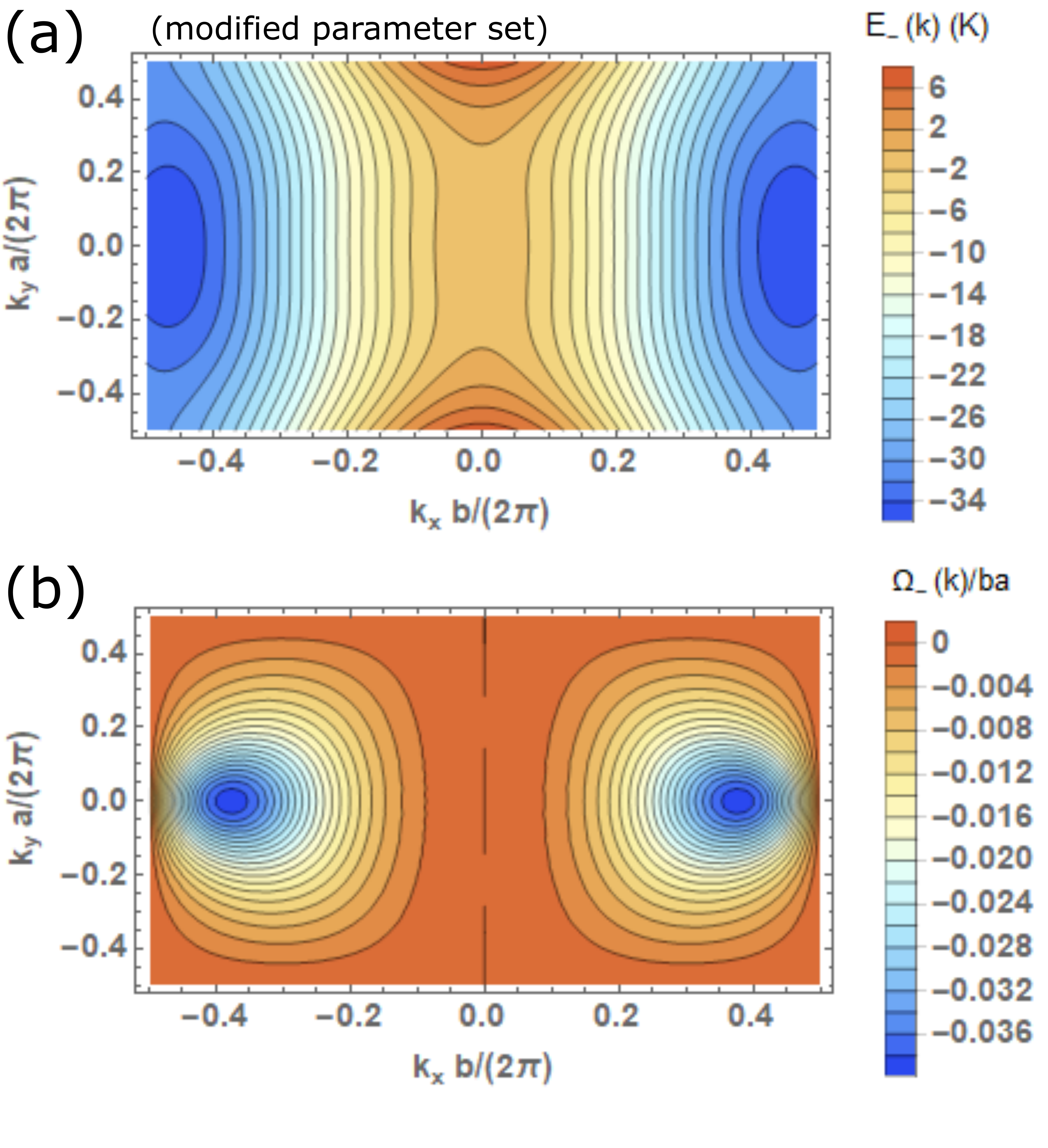}
\end{center}
\caption{\label{fig:B_dz_mod}
(Color online)
(a) The energy $E_-(\kv)$ [Eq.~\eqref{eq:epsilon_pm_k}] and
(b) the dimensionless Berry curvature $\Omega_- (\kv)/(ba)$ [Eqs.~\eqref{eq:Omega_psi} and \eqref{eq:Omega_theta_phi}]
for the lower magnon band over the first Brillouin zone.
These are calculated for the modified parameter set \eqref{eq:J_trial}.
}
\end{figure}
%############################

% [ Quantitative requirements ]--------------------------------------
We have seen that the magnetization and the thermal Hall conductivity calculated from our effective model
can capture some qualitative features of the experimental data \cite{Ishikawa15,Watanabe16},
but there are some quantitative discrepancies.
This indicates the necessity to modify the effective coupling constants, $\{\Jcal_{\Delta \rv,X}\}$ and $\Dcal_1$,
from the DFT-based parameter set in Table \ref{table:estimates} and Eq.\ \eqref{eq:Dcalv1}.
In tuning these coupling constants, the following observations provide a useful guidance.
Firstly, the magnetization data have indicated that the low-energy density of states of magnons, $G$,
must be roughly twice as large as the value for the DFT-based parameter set (see Fig.\ \ref{fig:magh}).
Secondly, the experimental data of $\kappa_{xy}$ for $H=15$ T \cite{Watanabe16} are
two orders of magnitude larger than our data for $H\approx 20$-$35$~T in Fig.~\ref{fig:kappaoT}. 
If we assume a smooth change of $\kappa_{xy}$ between these values of $H$,
the Berry curvature $\Omega_-(\kv)$ must be two orders of magnitude larger than our calculated data in Fig.~\ref{fig:B_dz}(b).
Furthermore, for the sign of $\kappa_{xy}$ to be consistent with experiment, 
$\Omega_-(\kv)$ must have the sign opposite to Fig.~\ref{fig:B_dz}(b). 
Lastly, the experimental value $H_\cone=27.5$ T \cite{Ishikawa15,Kohama19} of the low-field end of the $\frac13$-plateau poses another constraint.
We can therefore summarize the quantitative requirements for the effective model as follows:
\begin{subequations}\label{eq:constraints_J}
\begin{align}
 &h_\cone=(g\mu_B/k_B)\times 27.5~\mathrm{T}=39.7~\mathrm{K}, \label{eq:constraint_hc1}\\
 &G=\frac{1}{\pi\sqrt{C_xC_y}}=0.045~\mathrm{K}^{-1}, \label{eq:constraint_G}\\
 &\frac{\Omega_- (\kv_*)}{ba} \sim -0.03. \label{eq:constraint_Omega}
\end{align}
\end{subequations}
Here, for the Berry curvature $\Omega_-(\kv)$, we take the representative value at $\kv=\kv^*$,
for which a simple analytical expression \eqref{eq:Omega_kstar} is available.
The low-energy density of states, $G$, can be determined by the curvature
($C_x$ and $C_y$ in the $x$ and $y$ directions as in Eq.\ \eqref{eq:em_expand})
of the dispersion relation around the minima.
We note that for a given parameter set, accurate values of $h_\cone$, $C_x$, and $C_y$
can be obtained by numerically minimizing the dispersion relation in Eq.~\eqref{eq:epsilon_pm_k};
however, to see how they depend on the effective coupling constants,
the analytical (yet approximate) expressions in Eq.\ \eqref{eq:CxCy_est} in Appendix \ref{app:magdis} (valid for $2\Jcal_2>-\Jcal_1\gg$ others) are useful.

%############################
\begin{figure}
\begin{center}
\includegraphics[width=0.4\textwidth]{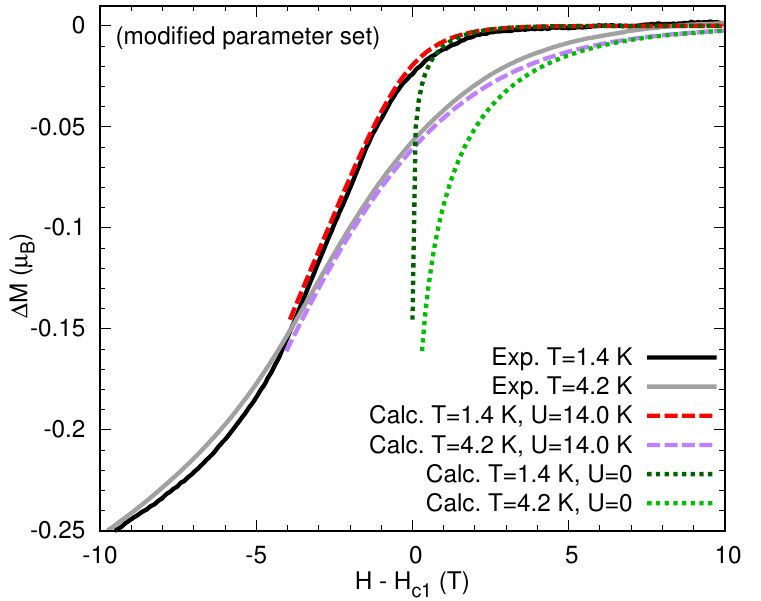}
\end{center}
\caption{\label{fig:magh_mod3}
(Color online)
Magnetization curves (measured from the $\frac13$-plateau) at $T=1.4$ and $4.2$ K.
The theoretical curves are calculated for the modified parameter set \eqref{eq:J_trial}.
For the effective interaction parameter $U=14.0$ K tuned appropriately [see Appendix \ref{app:EstU}],
the theoretical curves show a good agreement with the experimental data \cite{Ishikawa15}.
The curves for the non-interacting case $U=0$ are also presented for comparison.
}
\end{figure}
%############################

% [ Modified coupling constants ]--------------------------------------
To satisfy the requirements  in Eq.\ \eqref{eq:constraints_J},
we modify the coupling constants
%of the original model in Fig.~\ref{fig:volbeffDM}(a)
in the following way.
To enlarge $\Omega_-(\kv_*)/(ba)$ by two orders of magnitude, one must reduce the magnitude of the effective coupling $\Jcal_1$  considerably,
which is achieved by reducing $|J_1|$ or enlarging $|J'|$ from the DFT+$U$ estimates in Eq.~\eqref{eq:J_volb}.
Since the reduction in $|\Jcal_1|$ leads to enlargement in $h_\cone$,
we must reduce $J_2$ (and thus $\Jcal_2$) at the same time to satisfy Eq.\ \eqref{eq:constraint_hc1}.
We also introduce a modification $\delta\Jcal_3$ to the $\Jcal_3$ coupling as in Eq.\ \eqref{eq:dJcal3}
to change the sign of the Berry curvature to negative as required in Eq.\ \eqref{eq:constraint_Omega}.
After some examination, we have arrived at the following modified parameter set:
\begin{equation}\label{eq:J_mod}
 J:J':J_1:J_2=1:-0.5:-0.3:0.1,~~\delta\Jcal_3=6~\mathrm{K}.
\end{equation}
Here, the values of $J$, $g$, and the DM interactions are kept unchanged from Eqs.\ \eqref{eq:J_volb} and \eqref{eq:DvDv1}.
The effective coupling constants are then given by
\begin{equation}\label{eq:J_trial}
\begin{split}
&\Jcal_1=-7.2,~\Jcal_2=22.4,~\Jcal_2'=1.1,\\
&\Jcal_3=3.6,~\Jcal_3'=0.0,~\Jcal_4=-1.1,~\Jcal_5=-0.3,\\
&\Dcalv_1=(\Dcal_1^x,\Dcal_1^y,\Dcal_1^z)=(-1.3, 2.0, 1.9)~(\mathrm{K}),
\end{split}
\end{equation}
for which $h_\cone=35.9~\mathrm{K}=(g\mu_B/k_B)\times 24.8$~T, $(C_x,C_y)=(26.2,2.16)$~K,
$G=0.0423$~K$^{-1}$, and $\Omega_- (\kv_*)/(ba) =-0.0200$,
approximately satisfying Eq.\ \eqref{eq:constraints_J}.
We note that Eq.\ \eqref{eq:J_mod} is not a unique choice,
and there are various other ways of approximately satisfying Eq.\ \eqref{eq:constraints_J}.

%Jp := -0.5 J; J1 := -0.3 J; J2 := 0.1 J; dJe3 := 6;
%{Je1, Je2, Je2p, Je3, Je3p, Je4, Je5}
%={-7.19341, 22.3896, 1.14074, 3.61481, 0.0259259, -1.07852, -0.331852}
%Dve1={-1.33333, 2.00741, 1.91852}
%-Je2p + Je3 + Je3p=2.5
%{hs, hs kB/(g muB), 1/(Pi Sqrt[Cx Cy]), Brep, Q/(2 Pi), Cx, Cy}
%={35.8747, 24.8292, 0.0423467, -0.0199485, 0.473912, 26.1722, 2.15884}
%{hses, hses kB/(g muB), 1/(Pi Sqrt[Cxes Cyes]), Qes/(2 Pi), Cxes, Cyes}
%={35.414, 24.5104, 0.0418273, 0.480878, 21.9458, 2.63894}

%############################
\begin{figure}
\begin{center}
\includegraphics[width=0.5\textwidth]{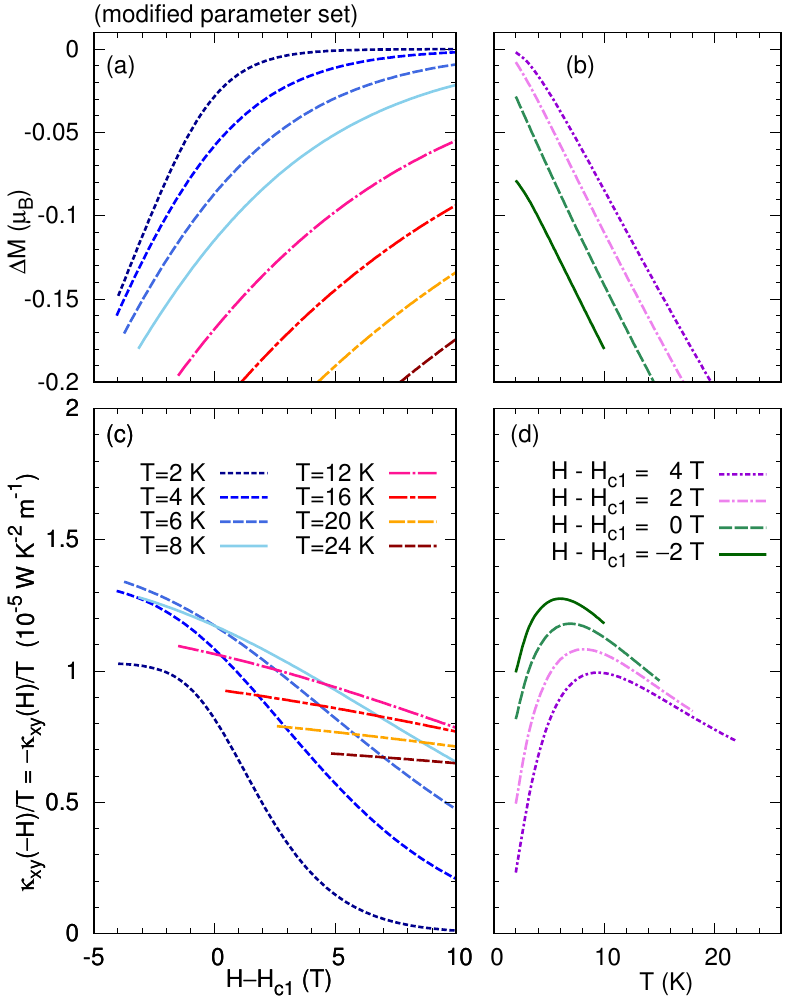}
\end{center}
\caption{\label{fig:kappaoT_mod}
(Color online)
(a,b) The magnetization (measured from the $\frac13$-plateau)
and (c,d) the thermal Hall conductivity $\kappa_{xy}(-H)/T=-\kappa_{xy}(H)/T$
as functions of $H-H_\cone$ and $T$  for the modified parameter set \eqref{eq:J_trial}.
}
\end{figure}
%############################

% [ Physical quantities for the modified coupling constants ]--------------------------------------
For the parameter set \eqref{eq:J_trial},
the lower magnon band $E_-(\kv)$, the dimensionless Berry curvature $\Omega_-(\kv_*)/(ba)$,
the magnetization, and the thermal Hall conductivity are presented in  Figs.\ \ref{fig:B_dz_mod}, \ref{fig:magh_mod3}, and \ref{fig:kappaoT_mod}.
The energy band $E_-(\kv)$  in Fig.\ \ref{fig:B_dz_mod}(a) shows minima at $\kv=\pm(Q/b,0)$ with $Q/(2\pi)=0.473$.
The Berry curvature $\Omega_-(\kv_*)/(ba)$ in Fig.\ \ref{fig:B_dz_mod}(b)  is negative over the entire Brillouin zone,
and have much larger amplitudes than in Fig.\ \ref{fig:B_dz}(b).
Reflecting this, the thermal Hall conductivity in Fig.\ \ref{fig:kappaoT_mod}(c,d) is positive
and have much larger amplitudes than in Fig.\ \ref{fig:kappaoT}(c,d).
Furthermore, the calculated magnetization curves show a good agreement with the experimental data as seen in Fig.\ \ref{fig:magh_mod3};
reflecting the larger magnon density of states $G$,
the magnon density $n\propto -\Delta M$  is larger in Figs.\ \ref{fig:magh_mod3} and \ref{fig:kappaoT_mod}(a,b) than in Figs.\ \ref{fig:magh} and \ref{fig:kappaoT}(a,b).

%================================================
\subsection{Other field directions} \label{sec:spwv_hxy}
%================================================

We have so far considered the case in which the field $\Hv$ is applied in the $-\hat{z}$ direction [Eq.\ \eqref{eq:H_z}].
Here we briefly discuss the case of other field directions.

When the field $\Hv$ is changed to the $-\hat{y}$ direction, the calculation of $\kappa^{xy}$
can be done in parallel with the case of $\Hv\parallel-\hat{z}$ discussed above
by just replacing $\Dcal_1^z$ by $\Dcal_1^y$.
Therefore, by measuring $\kappa^{xy}$ for two different directions of the field ($-\hat{y}$ and $-\hat{z}$),
one can determine the relative sign and magnitude of $\Dcal_1^y$ in comparison with $\Dcal_1^z$.

When the field $\Hv$ is changed to the $-\hat{x}$ direction, the system restores the $2_1$ screw axis symmetry.
In this case, we have $\kappa^{xy}=0$ because the heat current in the $y$ direction
changes its sign under the $2_1$ operation.

%%%%%%%%%%%%%%%%%%%%%%%%%%%%%%%%%%%%%%%%%%%%%%%%%
\section{Effective field theory for a quasi-one-dimensional regime} \label{sec:fieldtheory}
%%%%%%%%%%%%%%%%%%%%%%%%%%%%%%%%%%%%%%%%%%%%%%%%%

\newcommand{\conexy}{\mathrm{cone}}
\newcommand{\SDW}{\mathrm{SDW}}
\newcommand{\AFM}{\mathrm{AFM}}

% [ Section introduction ]--------------------------------------
In this section, we analyze the low-temperature phases of the effective pseudospin-$\frac12$ Hamiltonian \eqref{eq:Heff2} by means of a field-theoretical method.
The key point of our approach is the anisotropic triangular lattice structure of the effective model as seen in Fig.\ \ref{fig:volbeffDM}(b).
Among various effective couplings in Table \ref{table:estimates}, $\Jcal_2$ on the horizontal bonds has the largest magnitude.
If we look only at the $\Jcal_2$ coupling and the Zeeman term,
the system can be viewed as an array of decoupled Heisenberg chains in a magnetic field.
In this case, the low-energy description of each chain is given by the Tomonaga-Luttinger liquid (TLL) theory
unless the pseudospins are fully polarized by the field.
We can then include the other couplings in the effective Hamiltonian, which couple different chains and give rise to a variety of magnetic orders.
We analyze the competition among those inter-chain couplings through a perturbative renormalization group (RG) method and the chain mean field theory.

% [ Hamiltonian ]--------------------------------------
To facilitate the quasi-one-dimensional viewpoint, it is useful to represent the positions $\rv\in A\cup B$ using the $x$ and $y$ coordinates.
The $\Jcal_2$ and Zeeman parts of the effective Hamiltonian can then be rewritten as
\begin{equation}\label{eq:H2HZ}
 H_2 + H_\mathrm{Z} = \sum_y \left[ \Jcal_2\sum_x \Tv_{x,y}\cdot\Tv_{x+1,y} -\hv\cdot \sum_x \Tv_{x,y}\right].
\end{equation}
Here, the coordinates $y\in \mathbb{Z}$ and $x\in\mathbb{Z}+y/2$ are measured in units of $a/2$ and $b$, respectively.
Even (odd) $y$'s correspond to the sites in $A$ ($B$).
Interactions between nearest-neighbor chains are given by the $\Jcal_1$ and $\Jcal_4$ couplings and the effective DM interaction, which are expressed as
\begin{equation}\label{eq:H14DM}
\begin{split}
H_1&=\Jcal_1 \sum_y\sum_x \Tv_{x,y}\cdot \left( \Tv_{x+\frac12,y+1} + \Tv_{x-\frac12,y+1} \right) ,\\
H_4&=\Jcal_4 \sum_y\sum_x \Tv_{x,y}\cdot \left( \Tv_{x+\frac32,y+1} + \Tv_{x-\frac32,y+1} \right) ,\\
H_\DM&=\sum_y (-1)^y \sum_x \Big[ -\Dcalvb_1\cdot (\Tv_{x,y}\times\Tv_{x+\frac12,y+1})\\
 &~~~~~~~~~~~~~~~~~~~~~~~+ \Dcalv_1\cdot(\Tv_{x,y}\times\Tv_{x-\frac12,y+1}) \Big].
\end{split}
\end{equation}
Interactions between next-nearest-neighbor chains are given by the $\Jcal_2'$, $\Jcal_3$, and $\Jcal_3'$ couplings, which are combined into the form
\begin{equation}\label{eq:H2p33p}
\begin{split}
 &H_2'+H_3+H_3'\\
 &= \sum_{y:\mathrm{even}} \Tv_{x,y}\cdot \left( \Jcal_2' \Tv_{x,y+2}+\Jcal_3'\Tv_{x+1,y+2}+\Jcal_3\Tv_{x-1,y+2} \right)\\
 &~~~+\sum_{y:\mathrm{odd}} \Tv_{x,y}\cdot \left( \Jcal_2' \Tv_{x,y+2}+\Jcal_3\Tv_{x+1,y+2}+\Jcal_3'\Tv_{x-1,y+2} \right).
\end{split}
\end{equation}
Throughout this section, we do not consider the $\Jcal_5$ coupling as it only leads to a slight modification of the field-theoretical parameters
(such as the velocity $v$ and the compactification radius $R$) of each chain.

% [ Comments on our analysis ]--------------------------------------
Our analysis in this section is closely analogous to those in Refs.~\cite{Starykh07,Starykh10,ChenBalents13},
in which spatially anisotropic triangular antiferromagnets have been studied in relation with Cs$_2$CuCl$_4$.
Therefore, we take some notations similar to these references, refer to some of their results, and adapt them to the present problem.
Some doubt may be cast on the applicability of the coupled-chain approach to volborthite
since $\Jcal_1$ has a magnitude comparable to $\Jcal_2$ in the estimates in Table \ref{table:estimates}.
However, numerical studies have indicated that predictions of the coupled-chain approach
can qualitatively continue up to rather large values of inter-chain couplings.
Examples include the SDW and cone phases in spatially anisotropic triangular antiferromagnet \cite{ChenBalents13}
and the dimer and vector chiral phases in the $J_1$-$J_2$ XXZ chain \cite{White96,Nersesyan98,HikiharaKaburagi01,FSOF12,Agrapidis19}.
Furthermore, in the present problem, $|\Jcal_1|$ can potentially be modified into a much smaller value as we have discussed in Sec.\ \ref{sec:constraints_J}.

%================================================
\subsection{Field theory for a Heisenberg chain}
%================================================

% [ Spin chain Hamiltonian ]--------------------------------------
Here we briefly summarize the field-theoretical (bosonized) description of a spin-$\frac12$ antiferromagnetic Heisenberg chain in a magnetic field,
which corresponds to a part of Eq.~\eqref{eq:H2HZ} with fixed $y$.
The Hamiltonian is given by
\begin{equation}\label{eq:H1D}
 H_\mathrm{1D}= J\sum_x \Sv_x\cdot\Sv_{x+1}-h\sum_x S_x^z,~~J>0,
\end{equation}
where $x$ runs over integers or half-integers, and $\Sv_x$ is the spin-$\frac12$ operator at the position $x$.
The magnetic field $h$ is chosen to be in the $z$ direction.
The magnetization $m=(1/L)\sum_x \langle S_x^z \rangle=\langle S_x^z \rangle$, with $L$ being the number of spins in the chain,
is conserved in this Hamiltonian.

%The magnetization curve $M(h)$ in the groundstate can be obtained from the Bethe ansatz solution \cite{}.

% [  ]--------------------------------------
For any $m$ less than the saturation, i.e., $m\in (-1/2,1/2)$, the low-energy description of Eq.\ \eqref{eq:H1D} is given by the TLL theory with the Hamiltonian
\begin{equation}\label{eq:H_TLL}
 H_\mathrm{1D} = \frac{v}{2} \int dx \left[ (\partial_x\phi)^2+(\partial_x\theta)^2 \right].
\end{equation}
Here, the bosonic fields $\phi(x)$ and $\theta(x)$ satisfy the commutation relation $[\phi(x),\theta(x')]=i\Theta(x-x')$,
where $\Theta(x)$ is the Heaviside step function (with $\Theta(0)=1/2$).
The field $\phi(x)$ is compactified on a circle of radius $R$, and $\theta(x)$ is analogously compactified with the radius $1/(2\pi R)$;
the vertex operators appearing in Eq.\ \eqref{eq:Scalk} below are consistent with this compactification.
The velocity $v$ and the compactification radius $R$ are smooth functions of $m$.
The magnetization curve $m(h)$ and the dependences of $v/J$ and $R$ on $m$ can be determined
by numerically solving the integral equations obtained from the Bethe ansatz \cite{Bogoliubov86,Cabra98,Affleck99}.
%For the plots of $m(h)$, $v(m)/J$, and $\eta(m)=2\pi R(m)^2$, see, e.g., Figs.\ 8-10 of Ref.\ \cite{Affleck99}.
In particular, the exponent $\eta(m)=2\pi R(m)^2$ monotonically decreases from $\eta(0)=1$ to $\eta(1/2)=1/2$ with the increase in $m$; see Fig.\ \ref{fig:CMFT}(c) shown later.
%\tm{$\eta(m)=2\pi R(m)^2$? This is the first appearance of symbol $\eta$.}

At a fixed magnetization $m$, the low-energy fluctuations of spins occur around particular wave vectors $k$.
Such wave vectors are $k=0$ and $\pi\pm 2\delta$ with $\delta=\pi m$ for the ``longitudinal'' spin component $S_x^z$ along the field direction,
and $k=\pm2\delta$ and $\pi$ for the ``transverse'' spin components $S_x^\pm=S_x^x\pm iS_x^y$ perpendicular to the field.
The spin operators can thus be decomposed as
\begin{equation}\label{eq:SzSpm_bos}
\begin{split}
S_x^z &=m+\Scal_0^z(x)+  e^{i(-\pi+2\delta)x} \Scal_{-\pi+2\delta}^z (x)\\
 &~~~+e^{i(\pi-2\delta)x} \Scal_{\pi-2\delta}^z (x) , \\
S_x^\pm &= e^{i2\delta x}\Scal_{2\delta}^\pm (x)+e^{-i2\delta x}\Scal_{-2\delta}^\pm(x)+e^{\pm i\pi x}\Scal_\pi^\pm (x).
\end{split}
\end{equation}
Here, the operators $\Scal_k^\mu(x)~(\mu=z,\pm)$ describe slowly varying components, and are expressed in terms of the bosonic fields as
\begin{equation}\label{eq:Scalk}
\begin{split}
&\Scal_0^z(x)=\frac1{2\pi R} \partial_x \phi,~
\Scal_{\mp \pi\pm 2\delta}^z(x)=\mp \frac{A_1}{2i} e^{\pm i\phi/R},\\
&\Scal_{\pm 2\delta}^+(x)=\pm \frac{A_2}{2i} e^{i2\pi R\theta}e^{\pm i\phi/R},~
\Scal_{\pm 2\delta}^-(x)=\Scal_{\mp 2\delta}^+(x)^\dagger,\\
&\Scal_\pi^\pm (x)=A_3 e^{\pm i2\pi R\theta}.
\end{split}
\end{equation}
The coefficients  $A_1$, $A_2$, and $A_3$ are also smooth functions of $m$,
and have been determined numerically in Ref.\ \cite{Hikihara01}.
Among the operators in Eq.\ \eqref{eq:Scalk}, $\Scal_{\mp\pi\pm2\delta}^z$ and $S_\pi^\pm$
with comparatively smaller scaling dimensions $\Delta_\SDW=1/(2\eta)$ and $\Delta_\pm=\eta/2$,  respectively,
can play particularly important roles in the low-energy physics.
The other operators, $\Scal_0^z$ and $S_{2\delta}^\pm$, have scaling dimensions $1$ and $\eta/2+1/(2\eta)$, respectively.

%================================================
\subsection{Field theory for coupled chains}\label{sec:fieldtheory_cchain}
%================================================
\newcommand{\gammab}{\bar{\gamma}}
\newcommand{\Scalb}{\bar{\Scal}}

Let us now consider the coupled-chain problem with Eqs.\ \eqref{eq:H2HZ}, \eqref{eq:H14DM}, and \eqref{eq:H2p33p}.
Here we present the field-theoretical expressions of various inter-chain couplings,
and discuss their relevances to the low-energy physics on the basis of their scaling dimensions.
More detailed discussions on the competition among those couplings are given in Sec.~\ref{sec:CMFT}.

%++++++++++++++++++++++++++++++++++++++++++++++++
\subsubsection{Magnetic field in the $z$ direction}\label{sec:fieldz}
%++++++++++++++++++++++++++++++++++++++++++++++++

% [ Spin operators ]--------------------------------------
We first consider the case in which the field $\hv$ is applied in the $z$ direction.
Using Eq.\ \eqref{eq:SzSpm_bos} for each chain labeled by $y$, the pseudospin operators are expressed as
\begin{equation}\label{eq:T_bos}
\begin{split}
T_{x,y}^z &=m+\Scal_{y;0}^z(x)+  e^{i(-\pi+2\delta)x} \Scal_{y;-\pi+2\delta}^z (x)\\
 &~~~+e^{i(\pi-2\delta)x} \Scal_{y;\pi-2\delta}^z (x) , \\
T_{x,y}^\pm &=e^{i2\delta x}\Scal_{y;2\delta}^\pm (x)+e^{-i2\delta x}\Scal_{y;-2\delta}^\pm (x) + e^{\pm i\pi x}\Scal_{y;\pi}^\pm (x)
\end{split}
\end{equation}
with $m=(1/\Ntri)\sum_{x,y}\langle T_{x,y}^z\rangle$ and $\delta=\pi m$.
The operators $\Scal_{y;k}^\mu(x)~(\mu=z,\pm)$ can be expressed in terms of the bosonic fields $\phi_y(x)$ and $\theta_y(x)$ defined on each chain, as in Eq.~\eqref{eq:Scalk}.

% [ J1 and J4 couplings ]--------------------------------------
The $\Jcal_1$ and $\Jcal_4$ couplings between nearest-neighbor chains in Eq.\ \eqref{eq:H14DM} are then expressed as
\begin{equation}\label{eq:H14bos}
\begin{split}
&H_1+H_4\\
&=\sum_y \int dx \Big[ 2(\Jcal_1+\Jcal_4)\left( m^2+\Scal_{y;0}^z\Scal_{y+1;0}^z\right)\\
&~~~~~-\gamma_\SDW \left( \Scal_{y;-\pi+2\delta}^z\Scal_{y+1;\pi-2\delta}^z + \mathrm{h.c.}\right) \\
&~~~~~ -\gamma_\conexy \left( -i\Scal_{y;\pi}^+\partial_x\Scal_{y+1;\pi}^- + \mathrm{h.c.}\right)+\dots\Big]
\end{split}
\end{equation}
with
\begin{equation}
\begin{split}
 \gamma_\SDW &=2\left(-\Jcal_1\sin\delta+\Jcal_4\sin3\delta \right),\\
 \gamma_\conexy &=\left(-\Jcal_1+3\Jcal_4\right)/2.
\end{split}
\end{equation}
As discussed by Starykh {\it et al.}\ \cite{Starykh07,Starykh10,Starykh14} (and numerically demonstrated in Ref.~\cite{ChenBalents13}),
the interactions in Eq.~\eqref{eq:H14bos} lead to competition between SDW and cone orders.
Namely, the $\gamma_\SDW$ term with the scaling dimension $2\Delta_\SDW=1/\eta$ induces the incommensurate SDW order at low fields
while the $\gamma_\conexy$ term with the scaling dimension $1+2\Delta_\pm=1+\eta$ induces the incommensurate cone order at high fields (below the saturation);
see Fig.\ \ref{fig:CMFT}(c) for the plots of these scaling dimensions.
The interaction $\Scal_{y;0}^z\Scal_{y+1;0}^z$ with the scaling dimension $2$ is marginal for any $m\in (-1/2,1/2)$.
The ellipsis in Eq.~\eqref{eq:H14bos} indicates other terms which have larger scaling dimensions and are less important in the low-energy physics.

% [ J2', J3, J3', DM couplings ]--------------------------------------
The simple scenario of the SDW-cone competition in Eq.~\eqref{eq:H14bos} may break down if we also consider other couplings in the effective Hamiltonian. %\eqref{eq:Heff2}.
Specifically, the $\Jcal_2'$, $\Jcal_3$, and $\Jcal_3'$ couplings between next-nearest-neighbor chains in Eq.\ \eqref{eq:H2p33p} are expressed as
\begin{equation}\label{eq:H233bos}
\begin{split}
&H_2'+H_3+H_3'\\
&=\sum_y \int dx \Big\{ (\Jcal_2'+\Jcal_3+\Jcal_3')\left( m^2+\Scal_{y;0}^z\Scal_{y+2;0}^z\right)\\
&~~~~~-\left[ \left(\gamma_\SDW'+i(-1)^y\gamma_\SDW''\right) \Scal_{y;-\pi+2\delta}^z\Scal_{y+2;\pi-2\delta}^z +\mathrm{h.c.}\right]\\
&~~~~~-\frac{\gamma'}{2} \left(\Scal_{y;\pi}^+\Scal_{y+2;\pi}^-+\mathrm{h.c.} \right) \\
&~~~~~+(-1)^y\gamma_\conexy' \left(\Scal_{y;\pi}^+\partial_x\Scal_{y+2;\pi}^- + \mathrm{h.c.}\right)+\dots\Big\}
\end{split}
\end{equation}
with
\begin{equation}\label{eq:gammap}
\begin{split}
 \gamma_\SDW'&=-\Jcal_2'+(\Jcal_3+\Jcal_3')\cos 2\delta,\\
 \gamma_\SDW''&=(\Jcal_3-\Jcal_3')\sin 2\delta,\\
 \gamma'&=-\Jcal_2'+\Jcal_3+\Jcal_3',\\
 \gamma_\conexy' &= (\Jcal_3-\Jcal_3')/2.
\end{split}
\end{equation}
The $\gamma_\SDW'$ and $\gamma_\SDW''$ terms have the same scaling dimension as the $\gamma_\SDW$ term
while the $\gamma_\conexy'$ term has the same scaling dimension as the $\gamma_\conexy$ term.
The $\gamma'$ term has a smaller scaling dimension $2\Delta_\pm=\eta$, and grows much faster than the other terms in the RG flow;
when $\gamma'>0$ ($\gamma'<0$), it has the effect of stabilizing (destabilizing) the cone order induced by the $\gamma_\conexy$ term.
However, whether this term dominates the low-energy physics depends on the initial (bare) value of $\gamma'$, and this issue is analyzed in more detail in Sec.\ \ref{sec:CMFT}.

For $m\ne 0$, the effective DM interaction in Eq.\ \eqref{eq:H14DM} is expressed as
\begin{equation}\label{eq:DMz}
H_{\DM}=\Dcal_1^z \sum_y (-1)^y \int dx \left[ \frac12 \left( \Scal_{y;\pi}^+\partial_x\Scal_{y+1;\pi}^- + \mathrm{h.c.} \right) +\dots\right].
\end{equation}
We note that the $x$ and $y$ components of the DM interactions disappear in a perturbative treatment
because $T_{x,y}^\pm$ and $T_{x,y}^z$ have Fourier components with separated wave vectors for $m\ne 0$ as seen in Eq.~\eqref{eq:T_bos}.
The interaction in Eq.\ \eqref{eq:DMz} has the form similar to the $\gamma_\conexy$ term in Eq.~\eqref{eq:H14bos}.
In fact, the two terms can be combined as
\begin{equation}
\begin{split}
&\sum_y \int dx~i\left( \gamma_\conexy-i(-1)^y\frac{\Dcal_1^z}{2}\right)\Scal_{y;\pi}^+\partial_x\Scal_{y+1;\pi}^- + \mathrm{h.c.}\\
&=\sum_y \int dx~\gammab_\conexy \left( i\Scalb_{y;\pi}^+\partial_x\Scalb_{y+1;\pi}^- + \mathrm{h.c.}\right),
\end{split}
\end{equation}
where we define
\begin{equation}\label{eq:gauge_Scal}
\begin{split}
&\Scalb_{y;\pi}^\pm (x):= \exp\left[\pm i\frac{1-(-1)^y}{2} \nu\right] \Scal_{y;\pi}^\pm(x),\\
&\gammab_\conexy := \sqrt{\gamma_\conexy^2+(\Dcal_1^z/2)^2},\\
&\nu:=\arg \left( \gamma_\conexy + i\Dcal_1^z/2 \right).
%&\gamma_\conexy\pm i\Dcal_1^z/2=\gammab_\conexy e^{\pm i\mu},\\
\end{split}
\end{equation}
We note that the other terms in Eqs.~\eqref{eq:H14bos} and \eqref{eq:H233bos} remain unchanged under the ``gauge transformation'' of $\Scal_{y:\pi}^\pm(x)$ done here.
Therefore, the effects of the $\Dcal_1^z$ term are to enlarge the amplitude $\gammab_\SDW$ of the cone-inducing term
and to modify the resulting cone structure slightly via the gauge transformation.
The gauge transformation introduced here is analogous to the one discussed in the spin wave analysis; see the last paragraph of Sec.~\ref{sec:spwvH}.
Contrary to that case, the condition $\Jcal_4=0$ is not required in the present discussion.
This implies that in the low-energy theory, the thermal Hall conductivity $\kappa_{xy}$ vanishes even for $\Jcal_4\ne 0$.

%there, it was discussed that in the spin wave Hamiltonian with $\Jcal_4\ne 0$,
%the $\Dcal_1^z$ term can be absorbed into the $\Jcal_1$ term in terms of the transformed bosonic operators.
% in the present discussion, the effect of the $\Dcal_1^z$ term can be absorbed into the $\gammab$ term even in the

% [ m=0 case ]--------------------------------------
At sufficiently low fields, the $x$ and $y$ components of the DM interaction can also play certain roles in the low-energy physics
because the shift $\delta=\pi m$ of momenta in Eq.~\eqref{eq:T_bos} vanishes as $m\to 0$.
For a better understanding of this regime, it is useful to consider the case of precisely zero field, i.e., the case of $m=0$, as we do next.

%++++++++++++++++++++++++++++++++++++++++++++++++
\subsubsection{Zero magnetic field}
%++++++++++++++++++++++++++++++++++++++++++++++++

\begin{figure}
\begin{center}
\includegraphics[width=0.4\textwidth]{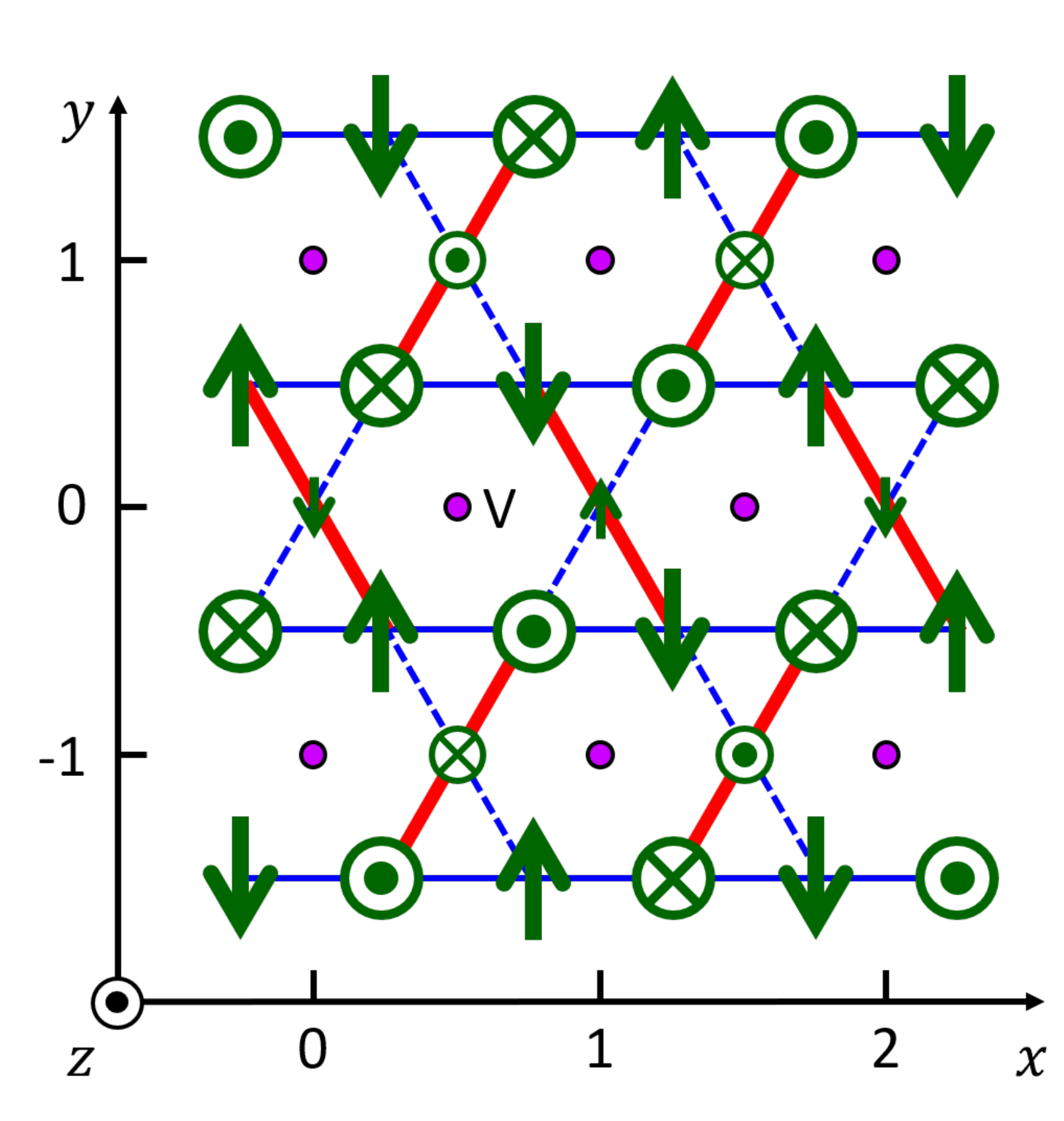}
\end{center}
\caption{\label{fig:orth}
(Color online)
Orthogonal order in the original spin-$\frac12$ system, based on Eqs.\ \eqref{eq:S_T} and \eqref{eq:orth_order} with $\Theta_0=0$.
This order is induced by the $\Dcal_1^x$ interaction in Eq.~\eqref{eq:HDM_h0} and formed in the $yz$ plane.
When a magnetic field is applied in the $x$ direction, the spins further acquire nonzero averages 
$\langle S_{\rv,1}^x\rangle=\langle S_{\rv,3}^x\rangle\approx\frac23 m$
and $\langle S_{\rv,2}^x\rangle \approx-\frac13 m$ in the $x$ direction, leading to a canted order.
In $^{51}$V NMR measurements as were done in Refs.\ \cite{Ishikawa15,YoshidaM17},
this order would show no direct signal because the internal fields at the V site from the surrounding spins on the hexagon
cancel out except the uniform component in the $x$ direction.
}
\end{figure}

For $m=0$,  Eq.~\eqref{eq:T_bos} can simply be written as
\begin{equation}
 \Tv_{x,y}=\Mv_y(x)+(-1)^{x-y/2}\Nv_y(x),
\end{equation}
which is based on the following mapping:
\begin{equation}
\begin{split}
 &\Scal_{y;0}^z \to M_y^z,\\
 &e^{-i\frac{\pi}{2} y}\Scal_{y;-\pi+2\delta}^z+e^{i\frac{\pi}{2} y}\Scal_{y;\pi-2\delta}^z\to N_y^z,\\
 &\Scal_{y;2\delta}^\pm+S_{y;-2\delta}^\pm \to M_y^\pm\equiv M_y^x\pm iM_y^y,\\
 &e^{\pm i\frac{\pi}{2}y}\Scal_{y;\pi}^\pm \to N_y^\pm\equiv N_y^x\pm iN_y^y.
\end{split}
\end{equation}
Here, the uniform and staggered components, $\Mv_y(x)$ and $\Nv_y(x)$, have the scaling dimensions $1$ and $1/2$, respectively.
The effective DM interactions can then be expressed as
\begin{equation}\label{eq:HDM_h0}
\begin{split}
 H_\DM=&\sum_y (-1)^y \int dx \sum_{b,c=x,y,z} \\
 \bigg[ &-\epsilon^{xbc} \Dcal_1^x \left( M_y^b \partial_x M_{y+1}^c+2N_y^bN_{y+1}^c +\dots \right)\\
 &+\sum_{a=y,z} \epsilon^{abc} \Dcal_1^a \left( 2M_y^bM_{y+1}^c+N_y^b\partial_xN_{y+1}^c +\dots \right) \bigg].
\end{split}
\end{equation}
The term $N_y^b N_{y+1}^c$ has the smallest scaling dimension $1$, and grows fastest in the RG flow.
If this term dominates the low-energy physics, an ``orthogonal'' order in which spins rotate by $\pm 90^\circ$ in the $yz$ plane (Fig.\ \ref{fig:orth}) appears.
Once this order appears at zero field, it is expected to persist in the low-field regime.

%############################
%############################

%++++++++++++++++++++++++++++++++++++++++++++++++
\subsubsection{Magnetic field in the $x$ direction}
%++++++++++++++++++++++++++++++++++++++++++++++++

% [ Field // x ]--------------------------------------
When the field is applied in the $x$ direction, we can perform the same line of analysis as in Sec.~\ref{sec:fieldz}
by expressing $T^x_{x,y}$ and $\tilde{T}^\pm_{x,y}:=T^y_{x,y}\pm iT^z_{x,y}$ in terms of $\Scalt^z_{y;k}(x)$ and $\Scalt^\pm_{y;k}(x)$, respectively,
in a way analogous to Eq.~\eqref{eq:T_bos}. %\tm{what do the tiled symbols mean?}
The only difference occurs in the expression of the effective DM interaction $H_\DM$.
Specifically, for $m:=\langle T^x_{x,y}\rangle\ne 0$, only the $x$ component of the DM interaction remains in a perturbative treatment,
and it is expressed as
\begin{equation}\label{eq:H_DM_x}
H_\DM
=-\Dcal_1^x \sum_y (-1)^y \int dx \Big[ \left( \Scalt_{y;\pi}^+\Scalt_{y+1;\pi}^- + \mathrm{h.c.} \right) +\dots\Big].
\end{equation}
This interaction is a finite-field version of the $N_y^bN_{y+1}^c$ term in Eq.\ \eqref{eq:HDM_h0}, and has a small scaling dimension $2\Delta_\pm=\eta$.
It thus grows as fast as the $\gamma'$ term in Eq.\ \eqref{eq:H233bos} along the RG flow,
and potentially dominates the low-energy physics over the entire range of the magnetic field below the $\frac13$-plateau.
If this happens, a {\it canted} orthogonal order in which pseudospins $\langle\Tv_{x,y}\rangle$ have the constant magnetization $m$ in the $x$ direction
and rotate by $\pm$90$^\circ$ in the $yz$ plane (as in Eq.\ \eqref{eq:orth_order} below) appears up to the $\frac13$-plateau.
Unfortunately, in $^{51}$V NMR measurements as were done in Refs.\ \cite{Ishikawa15,YoshidaM17},
this order would show no direct signal because of the cancellation of the internal fields at the V site as seen in Fig.\ \ref{fig:orth}.

%================================================
\subsection{Chain mean field theory}\label{sec:CMFT}
%================================================
\newcommand{\psit}{\tilde{\psi}}
\newcommand{\gammat}{\tilde{\gamma}}
\newcommand{\thetat}{\tilde{\theta}}
\newcommand{\thetab}{\bar{\theta}}
\newcommand{\mf}{\mathrm{mf}}
\newcommand{\orth}{\mathrm{orth}}

We now quantitatively analyze the competition among the inter-chain couplings which are described in Sec.~\ref{sec:fieldtheory_cchain}.
Specifically, following Ref.\ \cite{Starykh10}, we calculate the critical temperatures associated with different magnetic orders using the chain mean field theory.
The order with the highest critical temperature is expected to be selected among the competition.
We first summarize our results in Sec.~\ref{sec:CMFT_res}, and then describe the details of the analysis in the subsequent sections.
The processes of calculations go essentially the same way as in Appendix D of Ref.\ \cite{Starykh10},
and we roughly describe the ideas and adapt their results to the present model.

%############################
\begin{figure}
\begin{center}
\includegraphics[width=0.5\textwidth]{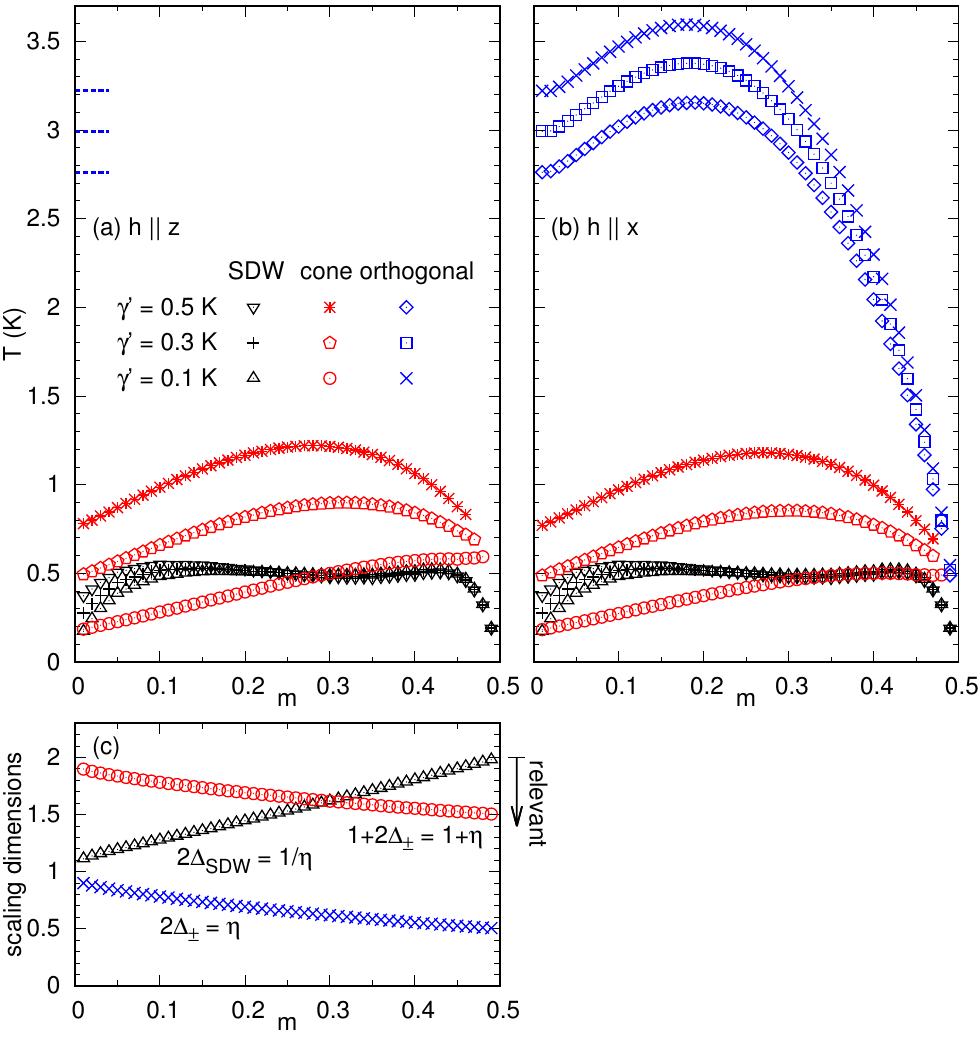}
\end{center}
\caption{\label{fig:CMFT}
(Color online)
Critical temperatures associated with the SDW, incommensurate cone, and canted orthogonal orders,
determined by the chain mean field theory for (a) $\hv\parallel \hat{z}$ and (b) $\hv\parallel \hat{x}$.
We use the modified parameter set in Eqs.\ \eqref{eq:J_mod} and \eqref{eq:J_trial},
but change the value of $\delta\Jcal_3$ to $\delta\Jcal_3\in\{3.6,3.8,4.0\}$ K, which correspond to $\gamma'\in \{0.1,0.3,0.5\}$ K, respectively.
In (a), the critical temperature for the orthogonal order (dashed horizontal lines at $T\sim 3$ K) is estimated from the data for smallest $m$ in (b);
the stability of this order against the change in $m$ is unfortunately beyond the scope of the present analysis.
(c) Scaling dimensions of the inter-chain couplings that induce the three types of orders.
The $\gamma_\SDW$ and $\gamma_\conexy$ terms in Eq.\ \eqref{eq:H14bos} have the scaling dimensions $2\Delta_\SDW$ and $1+2\Delta_\pm$, respectively,
whose crossing at $m\approx 0.3$ leads to the scenario of the SDW-cone competition.
The $\gamma'$ and $\Dcal_1^x$ terms in Eqs.\ \eqref{eq:H233bos} and \eqref{eq:H_DM_x} have the smaller scaling dimension $2\Delta_\pm$,
and substantially change the magnetic phase diagrams unless their magnitudes are suppressed.
}
\end{figure}
%############################

%------------------------------------------------
\subsubsection{Summary of the results}\label{sec:CMFT_res}
%------------------------------------------------

% [ Model parameter ]--------------------------------------
We consider the problem using the modified parameter set in Eqs.\ \eqref{eq:J_mod} and \eqref{eq:J_trial}.
If we directly use this parameter set, we have a rather large value ($\gamma'=2.5$ K) of the relevant $\gamma'$ coupling,
and the $\gamma'$ term dominates the low-energy physics over the entire range of the magnetic field below  the $\frac13$-plateau for $\hv\parallel \hat{z}$.
To obtain a rich phase diagram as observed experimentally \cite{Ishikawa15,YoshidaM17,Kohama19} (and shown in Fig.\ \ref{fig:phasemag}), the value of $\gamma'$ hence needs to be suppressed.
In our results presented here, we set $\delta\Jcal_3\in\{3.6,3.8,4.0\}$ K (instead of $\delta\Jcal_3=6.0$ K in Eq.\ \eqref{eq:J_mod}),
which correspond to $\gamma'\in \{0.1,0.3,0.5\}$ K, respectively.

% [ Results ]--------------------------------------
The calculated critical temperatures are displayed in Fig.~\ref{fig:CMFT}(a,b);
see also Fig.\ \ref{fig:phasemag} for the resulting phase diagrams at zero temperature.
We first look at the result for (a) $\hv\parallel\hat{z}$.
For $\gamma'=0.1$ K, which is sufficiently small, we find that the scenario of the SDW-cone competition in Eq.~\eqref{eq:H14bos} essentially holds:
as we lower the temperature, the SDW and cone orders first set in for $m\lesssim 0.3$ and $m\gtrsim 0.3$, respectively.
For $\gamma'=0.3$ and $0.5$ K, in contrast, the cone order is stabilized significantly by the $\gamma'$ coupling,
and it wins against the SDW order over the full range of the magnetization $m$.
At sufficiently small $m$, the orthogonal order induced by the $\Dcal_1^x$ interaction sets in
even before the cone order does, as indicated by dashed horizontal lines at $T\sim 3$ K in Fig.~\ref{fig:CMFT}(a).
We also note that close to the saturation $m=1/2$ of the pseudospin-$\frac12$ model, the bond nematic order appears due to condensation of bimagnons for a certain range of parameters \cite{Janson16};
unfortunately, we are not aware of an appropriate method for describing this order within bosonization for the present anisotropic triangular system.

Our result indicates that in order to obtain the SDW order over an extended range of the field as observed experimentally \cite{Ishikawa15,YoshidaM17},
the following constraint is required on the value of $\gamma'$:
\begin{equation}\label{eq:constraint_gammap}
 |\gamma'|=|-\Jcal_2'+\Jcal_3+\Jcal_3'|\lesssim 0.1~\mathrm{K}.
\end{equation}
Unfortunately, we have not been able to find a parameter set which simultaneously satisfies the requirements in Eqs.\ \eqref{eq:constraints_J} and \eqref{eq:constraint_gammap}.
We further note that the range of $\gamma'$ in Eq.\ \eqref{eq:constraint_gammap} is too narrow to be satisfied in a realistic system.
Since our analysis is performed in the limit of weak inter-chain couplings, we expect that
the constraint in Eq.\ \eqref{eq:constraint_gammap} is loosened with the increase in the magnitudes of the inter-chain couplings
in order to be consistent with experiment.
This indicates that a nontrivial stabilization mechanism of the SDW order exists beyond the scope of the perturbative RG approach.
We also note that defects present in volborthite crystals \cite{Hiroi19} can have nontrivial effects on the stabilization
of the SDW order---as the SDW state breaks only the translational symmetry, such defects can act as random fields
on the SDW order parameter, as is known in the context of collective pinning  \cite{Fukuyama78} (for further discussion, see Sec.\ V A of Ref.\ \cite{Starykh14}).

%We also note that defects present in volborthite crystals \cite{Hiroi19} can break chains into finite segments, leading to the Friedel oscillation of the local magnetizations \cite{Eggert95};
%this can strengthen the tendency towards the SDW order. \blue{More detailed consideration needed \cite{Starykh14,Fukuyama78}}

% [  ]--------------------------------------
We next look at the result for (b) $\hv\parallel\hat{x}$.
In this case, the canted orthogonal order first sets in owing to relevant $\Dcal_1^x$ over the full range of $m$ below the saturation. 

% [  ]--------------------------------------
Our results indicate that the magnetic phase diagrams depend sensitively on the direction of the magnetic field, as summarized in Fig.\ \ref{fig:phasemag}(b,c).
The experimental investigations of the phase diagram \cite{Ishikawa15,YoshidaM17,Kohama19} have mainly been conducted for magnetic fields perpendicular to the kagome plane, i.e., $\hv\parallel \hat{z}$.
It will be interesting if these experiments are extended to other directions of the magnetic field
to uncover nontrivial roles of the DM interactions as predicted here.

%------------------------------------------------
\subsubsection{SDW phase}\label{sec:SDW}
%------------------------------------------------

The inter-chain couplings related to a magnetic ordering in the longitudinal component $\langle T_{x,y}^z\rangle$ are summarized as
\begin{equation}\label{eq:HSDW}
\begin{split}
H_\SDW = \sum_y \int dx  &\frac{A_1^2}{2} \bigg[ -\gamma_\SDW \cos\frac{\phi_y-\phi_{y+1}}{R} \\
 &-\gamma_\SDW' \cos\frac{\phi_y-\phi_{y+2}}{R}\\
 &+ (-1)^y \gamma_\SDW'' \sin\frac{\phi_y-\phi_{y+2}}{R}
 \bigg].
\end{split}
\end{equation}
For $\gamma_\SDW>0$ and $\gamma_\SDW'=\gamma_\SDW''=0$, the ground state of this Hamiltonian is clearly given by $\phi_y(x)=\phi_0$ (constant).
Small $\gamma_\SDW'$ and $\gamma_\SDW''$ would not modify this ground state
since the expansion of \eqref{eq:HSDW} around this state does not produce any term linear in $\phi_y$'s.
By setting $\langle e^{i\phi_y/R}\rangle = \psit=|\psit|e^{i\Phi_0}$ and performing the mean-field decoupling of the inter-chain couplings, we obtain
\begin{equation}\label{eq:HSDW_mf}
 H_\SDW^\mf = - A_1^2 \left( \gamma_\SDW+\gamma_\SDW' \right) |\psit| \sum_y \int dx  \cos\left(\frac{\phi_y}{R}-\Phi_0 \right)  .
\end{equation}
The resulting state is the incommensurate SDW order in the longitudinal component along the magnetic field:
\begin{equation}
 \langle T_{x,y}^z\rangle = m+A_1|\psit| \sin\left[(\pi-2\delta)x-\Phi_0 \right].
\end{equation}
Combining Eq.\ \eqref{eq:HSDW_mf} with the unperturbed Hamiltonian, which is the TLL theory \eqref{eq:H_TLL} for each chain,
we obtain a collection of sine-Gordon models decoupled into different chains.
We can then calculate the finite-temperature average $\langle e^{i\phi_y/R}\rangle$ perturbatively in powers of $\psit$,
obtaining the self-consistent equation. The leading-order result is
\begin{equation}\label{eq:SC_SDW}
 |\psit|=A_1^2(\gamma_\SDW+\gamma_\SDW') |\psit| \chi_{\Delta_\SDW}(q=0,\omega_n=0;T)+\dots,
\end{equation}
where $\chi_{\Delta}$ is the momentum- and frequency-dependent susceptibility of the vertex operator
${\cal O}_\Delta=\cos \left(\sqrt{4\pi\Delta}\phi\right)$ [or ${\cal O}_\Delta=\cos\left(\sqrt{4\pi\Delta}\theta\right)$, which leads to the identical result],
evaluated in the TLL theory.
The condition that Eq.\ \eqref{eq:SC_SDW} acquires a nontrivial solution $|\psit|>0$
results in the following implicit equation for the critical temperature $T_\SDW$:
\begin{equation}
 1=A_1^2(\gamma_\SDW+\gamma_\SDW') \chi_{\Delta_\SDW}(q=0,\omega_n=0;T_\SDW).
\end{equation}
By solving this equation, the critical temperature is given by (see Eq.\ (D8) in Ref.~\cite{Starykh10})
\begin{equation*}
\begin{split}
 \left( \frac{2\pi T_\SDW}{v}\right)^{2-2\Delta_\SDW}
= \lambda_\SDW \frac{\Gamma(1-\Delta_\SDW)\Gamma^2(\Delta_\SDW/2)}{\Gamma(\Delta_\SDW)\Gamma^2(1-\Delta_\SDW/2)}&\\
\times\left[ 1+\lambda_\SDW \frac{\Gamma(\Delta_\SDW-1/2)}{\sqrt{\pi}(1-\Delta_\SDW)\Gamma(\Delta_\SDW)} \right]^{-1}&
\end{split}
\end{equation*}
with $\lambda_\SDW=\pi A_1^2\left( \gamma_\SDW+\gamma_\SDW' \right)/(2v)$.

%\begin{equation}
%\begin{split}
% H_\conexy = \sum_y \int dx A_3^2 \bigg\{
% &2\pi R\left( \partial_x\theta_y+\partial_x\theta_{y+1} \right)
% \left[ \gamma_\conexy\cos\left[2\pi R \left(\theta_y-\theta_{y+1}\right) \right]+(-1)^y\frac{\Dcal_1^z}{2}\sin\left[2\pi R \left(\theta_y-\theta_{y+1}\right) \right]\right]\\
%&-2\gamma' \cos\left[2\pi R \left(\theta_y-\theta_{y+2}\right) \right]
% +(-1)^y 2\pi R\gamma_\conexy' \left( \partial_x\theta_y+\partial_x\theta_{y+2} \right) \sin\left[2\pi R \left(\theta_y-\theta_{y+2}\right) \right] \bigg\}
%\end{split}
%\end{equation}

%------------------------------------------------
\subsubsection{Cone phase}\label{sec:cone}
%------------------------------------------------

We next discuss magnetic orderings in the transverse components $\langle T_{x,y}^\pm\rangle$,
first focusing on the case of  $\hv\parallel\hat{z}$.
In this case, the effective DM interaction can be treated by the suitable gauge transformation of $\Scal_{y;\pi}^\pm(x)$ in Eq.\ \eqref{eq:gauge_Scal}.
Reflecting this, we introduce the shifted field $\thetab_y(x)$ as
\begin{equation}
2\pi R\thetab_y = 2\pi R\theta_y + \frac{1-(-1)^y}{2}\nu.
\end{equation}
The inter-chain couplings related to a magnetic ordering in the transverse component are then summarized as
%\begin{widetext}
\begin{equation}\label{eq:H_cone}
\begin{split}
 H_\conexy = \sum_y \int dx A_3^2 \big\{
 &2\pi R \gammab_\conexy \left( \partial_x\thetab_y+\partial_x\thetab_{y+1} \right) \\
 &~~\times\cos\left[2\pi R \left(\thetab_y-\thetab_{y+1}\right) \right]\\
&-\gamma' \cos\left[2\pi R \left(\thetab_y-\thetab_{y+2}\right) \right]\\
 &+(-1)^y 2\pi R\gamma_\conexy' \left( \partial_x\thetab_y+\partial_x\thetab_{y+2} \right) \\
 &~~\times\sin\left[2\pi R \left(\thetab_y-\thetab_{y+2}\right) \right] \big\}.
\end{split}
\end{equation}
%\end{widetext}
As first pointed out by Nersesyan {\it et al.} \cite{Nersesyan98}, the interaction like the $\gammab_\conexy$ term here leads to an incommensurate transverse order.
To describe such an order, it is useful to set $2\pi R\thetab_y=-q_0x+2\pi R\thetat_y$, where the first term on the r.h.s.\ describes the incommensurate rotation of spins and the second the slowly varying component.
The inter-chain couplings in Eq.~\eqref{eq:H_cone} are then rewritten as
\begin{equation*}
\begin{split}
 H_\conexy = \sum_y &\int dx A_3^2 \bigg\{
 -2q_0 \gammab_\conexy\cos\left[2\pi R \left(\thetat_y-\thetat_{y+1}\right) \right]\\
 &-\gamma' \cos\left[2\pi R \left(\thetat_y-\thetat_{y+2}\right) \right]\\
 &+2(-1)^y q_0\gamma_\conexy' \sin\left[2\pi R \left(\thetat_y-\thetat_{y+2}\right) \right] \bigg\}.
\end{split}
\end{equation*}
By further setting $\langle e^{i2\pi R\thetat_y}\rangle = \psit=|\psit|e^{i\Theta_0}$ and performing the mean-field decoupling, we obtain
\begin{equation}
\begin{split}
H_\conexy^\mf = &-2A_3^2\left(2q_0\gammab_\conexy + \gamma'\right) |\psit| \\
 &\times\sum_y \int dx \cos\left(2\pi R\thetat_y-\Theta_0 \right).
\end{split}
\end{equation}
The resulting state is the incommensurate transverse order (the cone order) with
\begin{equation}
 \langle T_{x,y}^\pm\rangle = A_3|\psit| \exp\bigg\{ \pm i \left[(\pi-q_0)x-\frac{1-(-1)^y}{2}\nu+\Theta_0 \right]\bigg\}.
\end{equation}
We can determine the associated critical temperature $T_\conexy$ in a way similar to Sec.\ \ref{sec:SDW}.
However, reflecting the transformation from $\thetab_y$ to $\thetat_y$,
the susceptibility should be evaluated at the wave vector $q_0$.
The condition for the critical temperature $T_\conexy$ is then given by
\begin{equation}
 1=2A_3^2 \left( 2q_0\gammab_\conexy+\gamma' \right) \chi_{\Delta_\pm} \left( q_0,0;T_\conexy \right).
\end{equation}
The wave vector $q_0$ is determined in such a way as to maximize $T_\conexy$.
In this way, a set of implicit equations for determining $T_\conexy$ and $q_0$ are obtained as (see Eq.\ (D23) in Ref.~\cite{Starykh10})
\begin{equation}\label{eq:Tcone_q0}
\begin{split}
4\mathrm{Im}~\Psi\left( \frac{\Delta_\pm}{2}+ir \right)
=& \frac{2\pi\sinh(2\pi r)}{\cosh(2\pi r)-\cos(\pi \Delta_\pm)} \\
&+ \frac{\lambda_\conexy s}{\lambda_\conexy sr+\lambda'},\\
\frac{s^{2-2\Delta_\pm}}{\lambda_\conexy sr+\lambda'}
=& \frac{\Gamma(1-\Delta_\pm)}{\Gamma(\Delta_\pm)} \bigg| \Gamma\left( \frac{\Delta_\pm}{2}+ir \right) \bigg|^4 \\
&\times\left[ \cosh(2\pi r)-\cos(\pi\Delta_\pm) \right],
\end{split}
\end{equation}
where we introduce
\begin{equation*}
\begin{split}
&s=\frac{2\pi T_\conexy}{v},~
r=\frac{vq_0}{4\pi T_\conexy},\\
&\lambda_\conexy = \frac{2A_3^2\gammab_\conexy}{\pi v},~
\lambda'=\frac{A_3^2\gamma'}{2\pi v}.
\end{split}
\end{equation*}

When $\Dcal_1^x$ is sufficiently weak, a similar cone order can also appear for $\hv\parallel\hat{x}$.
Such a case can be analyzed by setting $\Dcal_1^z\to 0$ in the above argument.

%------------------------------------------------
\subsubsection{Orthogonal phase}\label{sec:orth}
%------------------------------------------------

We consider the case of $\hv\parallel\hat{x}$, when the $\Dcal_1^x$ interaction \eqref{eq:H_DM_x} can play a significant role.
This interaction can be rewritten as
\begin{equation}
%\begin{split}
 H_\DM = -\Dcal_1^x \sum_y (-1)^y \int dx 2A_3^2 \cos\left[2\pi R(\theta_y-\theta_{y+1}) \right] +\dots
% &+\dots \big\}
%\end{split}
\end{equation}
For $\Dcal_1^x<0$ as in Eq.~\eqref{eq:J_trial}, the ground state of this coupling is given by the state with
$\cos\left[2\pi R(\theta_y-\theta_{y+1}) \right]=-(-1)^y$, i.e.,
\begin{equation}
 2\pi R\theta_y(x)=n\pi+\Theta_0~~(y=2n-1,2n;~n\in\mathbb{Z}),
\end{equation}
where $\Theta_0$ is a constant. To analyze this order, it is useful to introduce the shifted field $\thetab_{y}(x)$ via
\begin{equation}
 2\pi R\theta_y(x)=2\pi R\thetab_{y}(x)+n\pi.
\end{equation}
The inter-chain coupling related to this order is then summarized as
\begin{equation}
\begin{split}
 H_\orth=\sum_y \int dx A_3^2 \big\{ &2\Dcal_1^x \cos\left[2\pi R(\thetab_y-\thetab_{y+1})\right]\\
 &+\gamma'\cos\left[2\pi R(\thetab_y-\thetab_{y+2})\right] \big\} .
\end{split}
\end{equation}
Here, we did not include the $\gamma_\conexy$ and $\gamma_\conexy'$ terms as they vanish after the mean-field treatment.
By setting $\langle e^{i2\pi R\thetab_y}\rangle=\psit=|\psit|e^{i\Theta_0}$ and performing the mean-field decoupling, we obtain
\begin{equation*}
 H_\orth^\mathrm{mf}=-\sum_y \int dx 2A_3^2 \left(-2\Dcal_1^x-\gamma' \right) |\psit|\cos\left(2\pi R\thetab_y-\Theta_0 \right).
\end{equation*}
The resulting state is the commensurate transverse order (the canted orthogonal order) with
\begin{equation}\label{eq:orth_order}
 \langle \Tt_{x,y}^\pm\rangle =\langle T_{x,y}^y\pm i T_{x,y}^z\rangle
 = A_3|\psit| e^{ \pm i \left[ \pi(n+x)+\Theta_0 \right]}.
\end{equation}
The condition for the critical temperature $T_\orth$ is given by
\begin{equation*}
 1=2A_3^2\left(-2\Dcal_1^x-\gamma' \right)\chi_{\Delta_\pm} (q=0,\omega_n=0;T_\orth),
\end{equation*}
which is independent of $\Theta_0$.
By solving this equation, the transition temperature is obtained as
\begin{equation}
\frac{s^{2-2\Delta_\pm}}{\lambda_\orth}
= \frac{\Gamma(1-\Delta_\pm)}{\Gamma(\Delta_\pm)} \bigg| \Gamma\left( \frac{\Delta_\pm}{2}\right) \bigg|^4
\left[ 1-\cos(\pi\Delta_\pm) \right],
\end{equation}
where we introduce
\begin{equation*}
 s=\frac{2\pi T_\orth}{v},~
 \lambda_\orth=\frac{A_3^2}{2\pi v} (-2\Dcal_1^x-\gamma').
\end{equation*}
This is similar to the second equation in Eq.\ \eqref{eq:Tcone_q0} but with $r=0$ because of the commensurate nature.
We note that $\Theta_0$ in Eq.~\eqref{eq:orth_order} should in the end be fixed at a certain value
as the effective spin model \eqref{eq:Heff2} does not possess a spin rotational symmetry around any axis owing to the DM interactions;
unfortunately, the value of $\Theta_0$ cannot be determined by the present mean field approach.

%%%%%%%%%%%%%%%%%%%%%%%%%%%%%%%%%%%%%%%%%%%%%%%%%
\section{Summary and outlook} \label{sec:summary}
%%%%%%%%%%%%%%%%%%%%%%%%%%%%%%%%%%%%%%%%%%%%%%%%%

In this paper, on the basis of the coupled-trimer model of Ref.\ \cite{Janson16},
we have investigated the effects of DM interactions on the magnetic properties of volborthite.
By means of a strong-coupling expansion,
we have derived an effective pseudospin-$\frac12$ model on an anisotropic triangular lattice.
In the effective model, the magnetic anisotropy is characterized by a single effective DM vector $\Dcalv_1$ (in contrast to four vectors in the original model),
which leads to a significant simplification of our analysis.
We have performed a spin wave analysis starting from the $\frac13$-plateau state for the case of magnetic fields perpendicular to the kagome layer.
The magnon Bloch states have been found to acquire a nonzero Berry curvature, which gives rise to a thermal Hall effect.
Our magnon Bose gas theory can explain qualitative features of the magnetization and the thermal Hall conductivity measured experimentally.
Through a further quantitative comparison with experiment,
%By comparing the theory with experimental data of the magnetization and the thermal Hall conductivity,
we have derived some constraints on the effective model as in Eq.\ \eqref{eq:constraints_J}.
In particular, the requirement of enlarging the Berry curvature by two orders of magnitude
leads to a much smaller magnitude of the $\Jcal_1$ coupling, promoting a quasi-one-dimensional picture.
Based on this picture, we have analyzed magnetic orders at low temperatures using effective field theory.
The requirement that the SDW phase appear for an extended range of the magnetic field poses the constraint \eqref{eq:constraint_gammap}
on the magnitude of the relevant $\gamma'$ coupling between the second-neighbor chains.
Assuming this, we have predicted the magnetic phase diagrams as schematically shown in Fig.\ \ref{fig:phasemag}(b,c), which sensitively depend on the field direction.
Unfortunately, we have not been able to find a parameter set which simultaneously satisfy the constraints in Eqs.\ \eqref{eq:constraints_J} and \eqref{eq:constraint_gammap},
leaving open the issue of more precise determination of the microscopic spin model of volborthite.

Our analysis of the thermal Hall effect has been focused on the regime just below the $\frac13$-plateau,
where the system can be described as a low-density gas of magnons.
This approach is less effective with lowering the magnetic field
as the mutual interactions between magnons become more significant.
It is worth noting that the pseudospin-$\frac12$ effective model on an anisotropic triangular lattice has a structure similar to that of Cs$_2$CuCl$_4$,
and may support fractionalized excitations such as spinons, psinons, and antipsinons in such an intermediate-field regime \cite{Kohno07,Kohno09}.
It is an interesting theoretical challenge to calculate the thermal Hall conductivity based on those fractionalized excitations.
Such a calculation can be directly compared with the thermal Hall conductivity data up to 15 T of Watanabe {\it et al.}\ \cite{Watanabe16}.
If experimental measurements can be extended to higher fields,
it would provide an exciting possibility of observing the crossover from fractionalized excitations to magnons through transport properties.
It would also be interesting to investigate the role of magnon bound states, which appear below 1 K around the low-field end of the $\frac13$-plateau, on transport properties.

The experimental investigations of the magnetic phase diagram \cite{Ishikawa15,YoshidaM17,Kohama19,Yamashita20} have mostly been conducted
for the case of $\Hv\parallel z$ as shown in Fig.\ \ref{fig:phasemag}(a).
Furthermore, the nature of Phase I has yet to be explored in single crystals.
We expect that the prediction of a crucial dependence on the field direction and the characterization of different phases in Fig. \ref{fig:phasemag}(b,c) in this work stimulate further experimental studies.
The nature of the two-step transition to Phase I with decrease in temperature \cite{YoshidaH12,YoshidaM17,Kohama19,Yamashita20} also merits further investigation in both theory and experiment.

%Further experimental investigation is desired to reveal the nontrivial dependence of the magnetic phase diagrams on the field direction as predicted in Fig.\ \ref{fig:phasemag}(b,c).
%Therefore, the magnetic phase diagram for $\Hv\parallel\hat{x}$ and its consistency with our prediction in Fig.\ \ref{fig:phasemag}(c) have yet to be investigated
%The nature of the two-step transition to Phase I with decrease in temperature \cite{YoshidaH12,YoshidaM17,Kohama19} also merits further investigation in both theory and experiment.
%In view of the non-collinear nature of the orthogonal order in Fig.\ \ref{fig:orth}, the spin-chirality decoupling scenario \cite{Teitel83,Miyashita84}
%may provide a key to understand the two-step transition to Phase I observed experimentally \cite{YoshidaH12,YoshidaM17,Kohama19}.

\bigskip

% [ Acknowledgments ]--------------------
The authors thank Z.\ Hiroi, H.\ Ishikawa, M.\ Yamashita, and M.\ Yoshida for sharing their experimental results,
and O.\ Janson for providing information on the DM interactions and for a collaboration on a related work.
The authors also acknowledge stimulating discussions with O.\ Benton, A.\ Furusaki, S.\ C.\ Furuya, J.\ Romh\'anyi, and O.\ Starykh.
This work was supported by
KAKENHI Grant No.\ JP18K03446 and No.\ JP16K05425 from the Japan Society for the Promotion of Science,
Matsuo Foundation, and Keio Gijuku Academic Development Funds.

\appendix

%%%%%%%%%%%%%%%%%%%%%%%%%%%%%%%%%%%%%%%%%%%%%%%%%
\section{Symmetry consideration of DM interactions} \label{app:DM}
%%%%%%%%%%%%%%%%%%%%%%%%%%%%%%%%%%%%%%%%%%%%%%%%%

Here we discuss how the symmetry of the space group P2$_1$/a (No.\ 14) imposes constraints on the DM vectors
as shown in Fig.~\ref{fig:volbeffDM}.

% [ Inversion ]--------------------------------------
Firstly, there is an inversion center at the center of each trimer.
Under the inversion ${\cal I}$ about $\rv\in A$ as shown in  Fig.~\ref{fig:volbeffDM}(a), the DM interactions $\Dv$ and $\Dv_1$ are transformed as
\begin{equation*}
\begin{split}
 &\Dv\cdot\left( \Sv_{\rv,1}\times \Sv_{\rv,2} \right)\\
 &\longrightarrow \Dv\cdot\left( \Sv_{\rv,3} \times \Sv_{\rv,2} \right)
 =-\Dv \cdot\left( \Sv_{\rv,2}\times \Sv_{\rv,3} \right),\\
 &\Dv_1\cdot\left( \Sv_{\rv,3}\times \Sv_{\rv+\uv,3} \right)\\
 &\longrightarrow \Dv_1\cdot\left( \Sv_{\rv,1} \times \Sv_{\rv-\uv,1} \right)
 =-\Dv_1 \cdot\left( \Sv_{\rv-\uv,1}\times \Sv_{\rv,1} \right).
\end{split}
\end{equation*}
We therefore find the appearance of the DM interactions $-\Dv$ and $-\Dv_1$ on the respective bonds.

% [ Screw axis ]--------------------------------------
Secondly, a two-fold screw ($2_1$) axis runs along each $J_1$-$J_2$ chain.
Namely, there is a symmetry under the shift by $\bv/2$ followed by $\pi$ rotation about the axis.
When the $2_1$ axis is chosen on the line on which $\Sv_{\rv,3}$ with $\rv\in A$ lies,
the DM interactions $-\Dv$ and $\Dv_1$ are transformed as
\begin{equation*}
\begin{split}
 &\Dv\cdot\left( \Sv_{\rv,1}\times \Sv_{\rv,2} \right)\\
 &\longrightarrow \Dv\cdot \left(\bar{\Sv}_{\rv+\uv,1}\times \bar{\Sv}_{\rv+\uv,2}\right)
 =\Dvb\cdot \left(\Sv_{\rv+\uv,1}\times\Sv_{\rv+\uv,2} \right),\\
 &\Dv_1 \cdot\left( \Sv_{\rv,3}\times \Sv_{\rv+\uv,3} \right)\\
 &\longrightarrow \Dv_1\cdot \left(\bar{\Sv}_{\rv+\uv,3} \times \bar{\Sv}_{\rv+\bv,3}\right)
 =\Dvb_1 \cdot \left(\Sv_{\rv+\uv,3}\times\Sv_{\rv+\bv,3} \right).
\end{split}
\end{equation*}
In these ways, we have constraints on the relative signs of the DM vectors as in Fig.~\ref{fig:volbeffDM}(a).

% [ DMs in the effective model ]--------------------------------------
Similar symmetry consideration also applies to the DM vectors in the effective model.
It leads to the relative signs of the DM vectors on the $\Jcal_1$ bonds as shown in Fig.~\ref{fig:volbeffDM}(b).
Furthermore, one can show that the DM interactions on the $\Jcal_2$ bonds strictly vanish.
This is because such a DM interaction, if present, is mapped onto the DM interaction with the reversed sign on the same bond
under the site-centered inversion followed by translation.

%%%%%%%%%%%%%%%%%%%%%%%%%%%%%%%%%%%%%%%%%%%%%%%%%
\section{Magnon band touching} \label{app:band_touch}
%%%%%%%%%%%%%%%%%%%%%%%%%%%%%%%%%%%%%%%%%%%%%%%%%

Here we argue that the touching of the two magnon bands \eqref{eq:epsilon_pm_k} at the Brillouin zone boundary
can be understood as the Kramers degeneracy due to certain antiunitary symmetries
(see Refs.\ \cite{Young15,ChenY16,Furusaki17,YangBJ17} for related arguments for other space groups).

We place the origin of the coordinate $(x,y,z)$ at the center of a trimer of type $A$.
We introduce the $2_1$ screw axis operation $\Ccal_{2x}$ about the axis $(y,z)=(-a/4,0)$,
the inversion $\Ical$ about the origin, and time reversal $\Theta$.
Under these operations, the coordinate, the momentum, and the spins are transformed as follows:
\begin{equation*}
\begin{split}
 \Ccal_{2x} :~&\rv=(x,y,z)\to \rv'=\left(x+\frac{b}{2},-y-\frac{a}{2},-z\right),\\
 &\kv\to\bar{\kv},~\Sv_{\rv,j}\to\bar{\Sv}_{\rv',j};\\
 \Ical :~&\rv\to -\rv,~\kv\to -\kv,~\Sv_{\rv,j} \to \Sv_{-\rv,4-j};\\
 \Theta :~&\rv\to \rv,~\kv\to-\kv,~\Sv_{\rv,j} \to-\Sv_{\rv,j}.
\end{split}
\end{equation*}
We note that $\Ccal_{2x}$ and $\Ical$ are unitary while $\Theta$ is antiunitary.

In the absence of a magnetic field $\Hv$, the Hamiltonian has the symmetries under all of the three operations $\Ccal_{2x}$, $\Ical$, and $\Theta$.
In the presence of a magnetic field $\Hv$ in the $z$ direction,
the symmetries under $\Ccal_{2x}$ and $\Theta$ are lost while that under $\Ical$ is retained.
Yet, the Hamiltonian is still symmetric under the following product of operations:
\begin{equation*}
\begin{split}
 \Ccal_{2x}\Theta:~&\rv=(x,y,z)\to \rv'=\left(x+\frac{b}{2},-y-\frac{a}{2},-z\right),\\
 &\kv\to-\bar{\kv},~\Sv_{\rv,j}\to-\bar{\Sv}_{\rv',j}.
\end{split}
\end{equation*}
It is also useful to consider the following product, which also leaves the Hamiltonian invariant:
\begin{equation*}
\begin{split}
 \Ical \Ccal_{2x} \Theta:~&\rv=(x,y,z)\to -\rv'=\left(-x-\frac{b}{2},y+\frac{a}{2},z\right),\\
 &\kv\to \bar{\kv},~\Sv_{\rv,j}\to-\bar{\Sv}_{-\rv',4-j}.
\end{split}
\end{equation*}
Since $(\Ccal_{2x}\Theta)^2$ and $(\Ical \Ccal_{2x}\Theta)^2$ are equal to the translations by $\bv$ and $\av$, respectively,
we have $(\Ccal_{2x}\Theta)^2=e^{i\kv\cdot \bv}$ and $(\Ical\Ccal_{2x}\Theta)^2=e^{i\kv\cdot\av}$ in the subspace with the wave vector $\kv$.
Thus, the Kramers degeneracy due to the antiunitary symmetry $\Ccal_{2x}\Theta$ occurs
when $e^{i\kv\cdot\bv}\ne 1$ and $\kv$ is invariant under this operation---this explains the band touching for $k_x=\pi/b$.
Similarly, the Kramers degeneracy due to the antiunitary symmetry $\Ical\Ccal_{2x}\Theta$ occurs
when $e^{i\kv\cdot\av}\ne 1$ and $\kv$ is invariant under this operation, i.e., when $(\kv\cdot\av,\kv\cdot\cv)=(\pi,0)$ and $(\pi,\pi)$;
since inter-layer couplings are neglected in our present model, this degeneracy occurs for arbitrary $\kv\cdot\cv$,
explaining the band touching for $k_y=\pi/a$.

%%%%%%%%%%%%%%%%%%%%%%%%%%%%%%%%%%%%%%%%%%%%%%%%%
\section{Analytical expression of the magnon density and estimation of the interaction parameter}\label{app:EstU}
%%%%%%%%%%%%%%%%%%%%%%%%%%%%%%%%%%%%%%%%%%%%%%%%%

Here we derive an analytical expression of $n(\mu,T)$ for $-\mu\ll T$,
and use it to estimate the interaction parameter $U$ from the slope of the experimental magnetization curve slightly below $H=H_\cone$.

In the expression of $n(\mu,T)$ in Eq.~\eqref{eq:n_T_mu},
particularly large contributions arise from the vicinity of the minima of the lower energy band at $\kv=(\pm Q/b,0)$.
Around these minima, the lower band $E_-(\kv)$ is expanded as in Eq.\ \eqref{eq:em_expand},
which leads to a constant density of states $G=1/(\pi \sqrt{C_xC_y})$ in units of $\Ntri/2$ at low energies.

% The lower band is expanded around these points as
% \begin{equation}\label{eq:em_expand}
%  E_-(\kv)\approx E_0+\frac{C_x}2 \left(k_xb\mp Q\right)^2 + \frac{C_y}2  \left( k_y a \right)^2.
% \end{equation}
% This leads to a constant density of states $G=1/(\pi \sqrt{C_xC_y})$ in units of $\Ntri/2$ at low energies.
% For the second-order model (see Table \ref{table:estimates}) with the DM vectors \eqref{eq:Dcalv1}, we have $(C_x,C_y)=(42.8, 5.12)$~K and $G=0.0215$~K$^{-1}$.

When $-\mu\ll T$, we can approximate the Bose distribution function around the minima of the lower band as
\begin{equation}\label{eq:rhom_approx}
 \rho_-(\kv) \approx \frac{T}{\Kv^2/2-\mu}
\end{equation}
with $\Kv:= \left( \sqrt{C_x}(k_xb\mp Q),\sqrt{C_y}k_ya \right)$.
This approximation is valid for $\Kv^2/2-\mu \ll T$.
Since the major contribution to Eq.~\eqref{eq:n_T_mu} comes from small $\Kv$,
we can approximate Eq.~\eqref{eq:n_T_mu} by the integral of Eq.~\eqref{eq:rhom_approx} over $|\Kv|<\sqrt{2T}$, obtaining
\begin{equation*}
\begin{split}
 n (\mu,T) &\approx \frac{T}{\sqrt{C_xC_y}} \int_{|\Kv|<\sqrt{2T}}\frac{d^2\Kv}{(2\pi)^2} \frac{1}{\Kv^2/2-\mu}\\
 &= \frac{GT}{2} \int_0^{\sqrt{2T}} dK \frac{K}{K^2/2-\mu}
 %&=\frac{T}{2\pi C} \ln \left( K^2/2-\mu \right) \Big|_{K=0}^{\sqrt{2T}}
 \approx \frac{GT}{2} \ln \left( \frac{T}{-\mu} \right).
\end{split}
\end{equation*}
We therefore have the relation
\begin{equation}\label{eq:h_n_asymp_app}
 -\mu = h-h_\cone+2Un \approx T \exp\left( - \frac{2 n}{GT} \right).
\end{equation}
This relation indicates that for $n\gg GT/2$, we have $h-h_\cone\approx -2Un$ and thus
the right-hand side of Eq.~\eqref{eq:M_n} is given by
\begin{equation}
 -g\mu_B \frac{n}{3} \approx \frac{g\mu_B}{6U} (h-h_\cone) = \frac{(g\mu_B)^2}{6Uk_B} (H-H_\cone).
\end{equation}
Therefore, $U$ can be determined from the slope $dM/dh$ of the magnetization curve as
\begin{equation}
 U=\frac{(g\mu_B)^2}{6k_B} \left( \frac{dM}{dh} \right)^{-1}.
\end{equation}

In the experimental magnetization data for $T=1.4$ K \cite{Ishikawa15},
a nearly constant slope $d\tilde{M}/dH=[0.0371~({\rm T}^{-1})]\mu_B$ has been found
in the field range between 23.3 and 25.9~T, which is slightly below $H_\cone\simeq 27.5$~T.
From this slope, the effective interaction parameter is estimated as $U=14.0$~K.
We note that the relation \eqref{eq:h_n_asymp_app} is also used to estimate the density of states of magnons, $G$, from the experimental data;
see Eq.\ \eqref{eq:h_n_asymp} and Fig. \ref{fig:magh}(b).

%%%%%%%%%%%%%%%%%%%%%%%%%%%%%%%%%%%%%%%%%%%%%%%%%
\section{Expansion of the magnon dispersion relation}\label{app:magdis}
%%%%%%%%%%%%%%%%%%%%%%%%%%%%%%%%%%%%%%%%%%%%%%%%%

\newcommand{\est}{\mathrm{est}}

Here we derive analytical expressions of $h_\cone:=-E_0$, $Q/(2\pi)$, and $(C_x,C_y)$,
which are introduced in Sec.\ \ref{sec:Bloch}.
Although these constants can be calculated accurately by numerically minimizing Eq.\ \eqref{eq:epsilon_pm_k},
the analytical (yet approximate) expressions that we derive here can clarify how they depend on the effective coupling constants.

To perform an analytical calculation, we first note that $\Jcal_1$ and $\Jcal_2$ have
much larger magnitudes than the other effective couplings
(see Table \ref{table:estimates} and Eq.\ \eqref{eq:Dcalv1}).
We therefore treat the latter couplings perturbatively.
By ignoring terms of order $\left(\Dcal_1^z\right)^2/|\Jcal_1|$ and $\left(\Jcal_3-\Jcal_3'\right)^2/|\Jcal_1|$,
the lower energy band $E_-(\kv)$ in Eq.\ \eqref{eq:epsilon_pm_k} is approximated as
\begin{equation}\label{eq:Em_k_approx}
\begin{split}
 E_-(\kv)&\approx E(\kv)-|J^x(\kv)|\\
 &=-\Jcal + 2\Jcal_1 \cos\frac{k_xb}{2}\cos\frac{k_ya}{2} + \Jcal_2 \cos\left(k_xb\right)\\
 &~~~+\Jcal_2' \cos\left(k_ya\right) + \left(\Jcal_3+\Jcal_3'\right) \cos\left(k_xb\right) \cos\left(k_ya\right)\\
 &~~~+2\Jcal_4 \cos\frac{3k_xb}{2}\cos\frac{k_ya}{2} + \Jcal_5 \cos\left(2k_xb\right).
\end{split}
\end{equation}
Henceforth, we assume $0<-\Jcal_1<2\Jcal_2$ and $\Jcal_1^2>4\Jcal_2\Jcal_2'$.
When $\Jcal_3+\Jcal_3'=\Jcal_4=\Jcal_5=0$, $E_-(\kv)$ above is minimized at
$\kv=(\pm Q_0/b,0)$ with
\begin{equation}
 \cos \frac{Q_0}{2}=-\frac{\Jcal_1}{2\Jcal_2}\equiv \eta.
\end{equation}
When $\Jcal_3+\Jcal_3'$, $\Jcal_4$, and $\Jcal_5$ are finite but their magnitudes are sufficiently smaller than $|\Jcal_1|$,
we can expect that these minimum points change only perturbatively.
We can therefore search for the minima of Eq.\ \eqref{eq:Em_k_approx} by expanding it around $\kv=(\pm Q_0/b,0)$.
The resulting expression is Eq.\ \eqref{eq:em_expand} with $Q=Q_0+\delta Q$,
and the first-order perturbative estimates of $h_\cone=-E_0$, $\delta Q$, and $(C_x,C_y)$ are obtained as
\begin{equation}\label{eq:CxCy_est}
\begin{split}
h_\cone^\est
&= 2\Jcal_2(1-\eta)^2+2\left(\Jcal_3+\Jcal_3'\right)\left(1-\eta^2\right)\\
&~~~+4\Jcal_4\left(1+3\eta-4\eta^3 \right) + 8\Jcal_5 \left(\eta^2-\eta^4\right),\\
\delta Q^\est
&=\frac{2\left(\Jcal_3+\Jcal_3'\right)\eta+3\Jcal_4 \left(4\eta^2-1\right)+8\Jcal_5\left(2\eta^3-\eta\right)}{\Jcal_2 \sqrt{1-\eta^2}},\\
C_x^\est
&= \Jcal_2\left(1-\eta^2\right)+\left(\Jcal_3+\Jcal_3'\right)\left(1-2\eta^2\right)\\
&~~~+\frac{9\Jcal_4}{2}\left(3\eta-4\eta^3\right)+4\Jcal_5\left(1-8\eta^2+8\eta^4 \right),\\
C_y^\est
&= \Jcal_2\eta^2-\Jcal_2'+\left(\Jcal_3+\Jcal_3'\right)\left(1-2\eta^2\right)+\frac{\Jcal_4}{2}\left(3\eta-4\eta^3\right).
\end{split}
\end{equation}
For the second-order model (see Table \ref{table:estimates}), we have
$h_\cone^\est=34.3$~K, $Q^\est/(2\pi)=0.382$, $(C_x^\est,C_y^\est)=(41.2, 5.76)$~K;
these agree reasonably with the accurate values
$h_\cone=35.2$~K, $Q/(2\pi)=0.369$, $(C_x,C_y)=(42.8, 5.12)$~K given in Sec.\ \ref{sec:Bloch}.

%%%%%%%%%%%%%%%%%%%%%%%%%%%%%%%%%%%%%%%%%%%%%%%%%
\section{Expression of the Berry curvature} \label{app:Berry}
%%%%%%%%%%%%%%%%%%%%%%%%%%%%%%%%%%%%%%%%%%%%%%%%%

Using Eq.~\eqref{eq:psi_pm}, the Berry curvature $\Omega_\pm(\kv)$ defined in Eq.~\eqref{eq:Omega_psi} can be rewritten
in terms of angular variables $\theta(\kv)$ and $\phi(\kv)$ as
\begin{equation}\label{eq:Omega_theta_phi}
 \Omega_\pm (\kv)
 = \pm \frac12 \sum_{i,j} \epsilon_{ij} \left(\partial_i \phi\right) \left(\partial_j \theta\right) \sin\theta .
\end{equation}
The derivatives appearing in this expression can be expressed in terms of the vector $\Jv(\kv)$ in Eq.~\eqref{eq:epsilon_J_k} as
\begin{equation}
\begin{split}
 &\partial_i \phi = \frac{1}{2i} \frac{J^-}{J^+} \partial_i \left( \frac{J^+}{J^-}\right)
 =\frac{1}{2i} \left( \frac{\partial_iJ^+}{J^+} - \frac{\partial_iJ^-}{J^-} \right),\\
 &\left(\partial_j \theta\right) \sin\theta = -\partial_j \left( \frac{J^z}{J} \right),
\end{split}
\end{equation}
where $J^\pm (\kv):=J^x(\kv)\pm iJ^y(\kv)$.
These expressions are used to calculate $\Omega_-(\kv)$ numerically in Fig.~\ref{fig:B_dz}(b).
At $\kv=\kv_*$, we have
\begin{equation*}
\begin{split}
 J^\pm (\kv_*+d\kv)
 =& \sqrt{2}(\Jcal_1-\Jcal_4\pm i\Dcal_1^z)\\
 &-\frac{1}{\sqrt{2}} (\Jcal_1+3\Jcal_4\pm i\Dcal_1^z) dk_x b+\mathcal{O}\left( (d\kv)^2 \right) ,\\
 J^z(\kv_*+d\kv)=&(\Jcal_3-\Jcal_3') dk_y a+\mathcal{O} \left( (d\kv)^2 \right),
\end{split}
\end{equation*}
which lead to the simple expression of $\Omega_\pm(\kv_*)$ in Eq.~\eqref{eq:Omega_kstar}.

%%%%%%%%%%%%%%%%%%%%%%%%%%%%%%%%%%%%%%%%%%%%%%%%%
\bibliography{volbDM.bib}
%%%%%%%%%%%%%%%%%%%%%%%%%%%%%%%%%%%%%%%%%%%%%%%%%

\end{document}